\documentclass[%
reprint,
 superscriptaddress,
 amsmath,amssymb,
 aps,
pra,
floatfix,
noeprint
]{revtex4-2}

\usepackage{graphicx}
\usepackage{dcolumn}
\usepackage{bm}
\usepackage{amssymb}
\usepackage{epstopdf}
\usepackage{textcomp}
\usepackage{enumerate}
\usepackage{epsfig,amsmath}
\usepackage{subfigure}
\usepackage[colorlinks]{hyperref}
\usepackage{color}
\usepackage{float}
\usepackage{wasysym}
\usepackage{MnSymbol}
\usepackage{tabularx}
\usepackage{soul}
\usepackage{upgreek}

\begin{document}

\title{Spin and valley degrees of freedom in a bilayer graphene quantum point contact: Zeeman splitting and interaction effects 
} 

\author{Vanessa Gall}
\affiliation{\mbox{Institute for Quantum Materials and Technologies, Karlsruhe Institute of Technology, 76021 Karlsruhe, Germany}}
\affiliation{\mbox{Institut f\"ur Theorie der Kondensierten Materie, Karlsruhe Institute of Technology, 76128 Karlsruhe, Germany}}

\author{Rainer Kraft}
\affiliation{Physikalisches  Institut,  Karlsruhe  Institute  of  Technology,  76131  Karlsruhe,  Germany}

\author{Igor V. Gornyi}
\affiliation{\mbox{Institute for Quantum Materials and Technologies, Karlsruhe Institute of Technology, 76021 Karlsruhe, Germany}}
\affiliation{\mbox{Institut f\"ur Theorie der Kondensierten Materie, Karlsruhe Institute of Technology, 76128 Karlsruhe, Germany}}
\affiliation{Ioffe Institute, 194021 St.~Petersburg, Russia}

\author{Romain Danneau}
\affiliation{\mbox{Institute for Quantum Materials and Technologies, Karlsruhe Institute of Technology, 76021 Karlsruhe, Germany}}

\begin{abstract}	
We present a study on the lifting of degeneracy of the size-quantized energy levels in an electrostatically defined quantum point contact in bilayer graphene by the application of in-plane magnetic fields. We observe a Zeeman spin splitting of the first three subbands, characterized by effective Land\'{e} $g$-factors that are enhanced by confinement and interactions. In the gate-voltage dependence of the conductance, a shoulder-like feature below the lowest subband appears, which we identify as a $0.7$ anomaly stemming from the interaction-induced lifting of the band degeneracy. We employ a phenomenological model of the $0.7$ anomaly to the gate-defined channel in bilayer graphene subject to in-plane magnetic field. Based on the qualitative theoretical predictions for the conductance evolution with increasing magnetic field, we conclude that the assumption of an effective spontaneous spin splitting is capable of describing our observations, while the valley degree of freedom remains degenerate.   
\end{abstract}

\maketitle
\section{Introduction}
%

Exploiting the quantum degrees of freedom of charge carriers offers a potential route for designing new types of quantum electronic devices. While most studied systems involve the electron's spin degree of freedom aiming at spintronic applications \cite{Wolf2001,Awschalom2007}, more recently the additional valley isospin in a variety of materials has attracted a growing interest for use in valleytronics \cite{Schaibley2016}. However, irrespective of the system of choice, the implementation of spin- or valley-based functionalities into electronic devices requires a full control of the quantum state itself. A quantum point contact that confines charge carriers into one dimension \cite{vanHouten1992}, is one of the basic building blocks for efficient injection, control, and read-out measures.

Recently, we have reported \cite{Kraft2018b} on an electrostatically-induced quantum point contact (QPC) in bilayer graphene (BLG) \cite{Allen2012,Goossens2012,Kraft2018a,Overweg2018a,Overweg2018,Banszerus2018,Banszerus2020, Sakanashi2021, Terasawa2021}, \textit{i.\,e.}, a system with four-fold spin and valley degeneracy, where the constriction is realized by local band gap engineering with a  displacement field perpendicular to the BLG plane. We observed confinement with well-resolved conductance quantization in steps of $4\,e^2/h$ down to the lowest one-dimensional (1D) subband, as well as a peculiar valley subband splitting and merging of $K$ and $K'$ valleys from two non-adjacent subbands in an out-of-plane magnetic field (see also Ref. \cite{Overweg2018}).

In the present paper, we investigate the same system in an in-plane magnetic field. In this context, we became aware of the publication \cite{Banszerus2018} that reported on conductance measurements in a similar setup and found certain features additional to the expected conductance quantization. These features were attributed \cite{Banszerus2018} to the substrate-induced Kane-Mele spin-orbit coupling \cite{Kane2005} below the lowest plateau. 
Since the reported values of the spin-orbit coupling in monolayer graphene is of the order of $40\,\mu$eV \cite{Sichau2019} (corresponding to temperatures of the order of $0.5$\,K) and there is no clear mechanism that would lead to an enhancement of spin-orbit coupling by hexagonal boron nitride (hBN), we expect another mechanism behind such features. Here, we explore alternative possibilities for the explanation of the appearance of additional features in the conductance. 

A very natural guess is that the lifting of degeneracy is due to interaction effects. 
While renormalization-group studies show that the Coulomb interaction in clean graphene becomes marginally irrelevant \cite{Gonzalez2001}, BLG behaves more like a  typical two-dimensional (2D) electron gas. Non-perturbative approaches to the effects of long-range interactions show that graphene may feature interaction-induced instabilities. These effects are expected to be particularly important in very clean samples, at very low densities, and in high magnetic fields. Proposed theories include superconducting instabilities \cite{Gonzalez2001, Gonzalez2008, Wang2012, Vucicevic2012}, (anti-)ferromagnetic instabilities  \cite{Coey2002, Stauber2007, Lang2012}, excitonic instabilities \cite{Dillenschneider2008, Milovanovic2008, Nandkishore2010}, and whole lot of others \cite{Wang2008, Valenzuela2008, Jung2015, Guinea2009, Abanin2010, Berman2010, Wang2010}. For a summary or comparison see, e.g., Refs.~\cite{Kiesel2012, Kotov2012, Scherer2012, Platt2013, Roy2014, Throckmorton2014, Cvetkovic2012}.

\begin{figure*}[htp]
	\centering
	\includegraphics[width=0.9
	\textwidth]{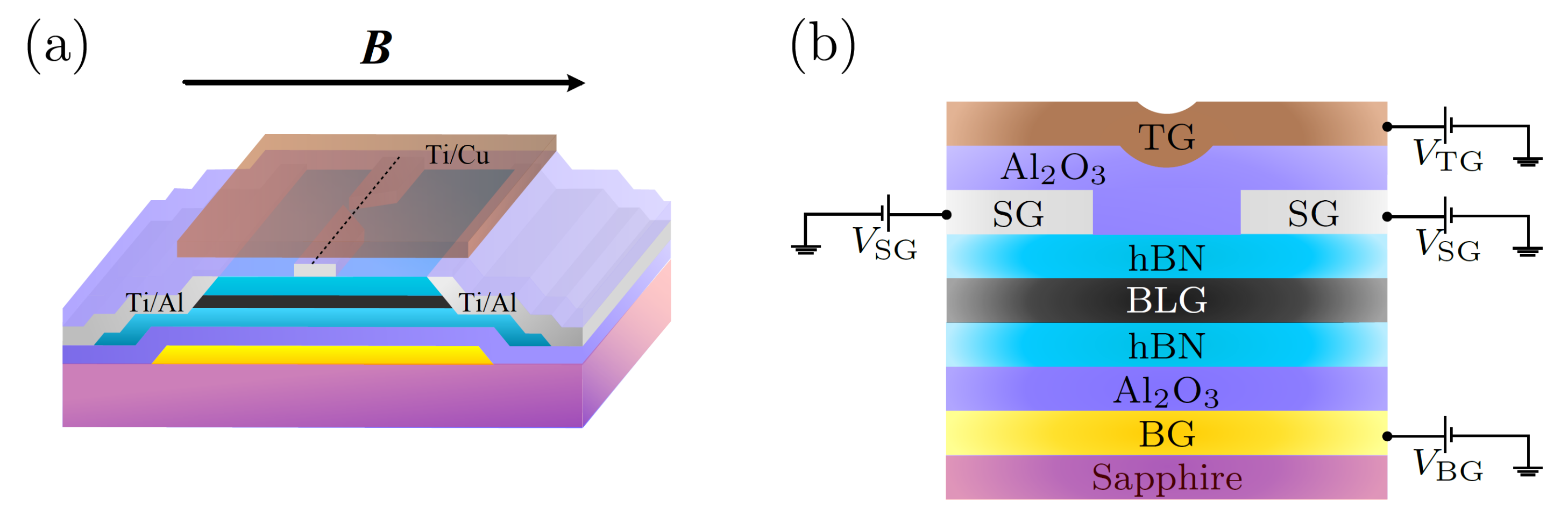}
	\caption{(a) Schematic illustration of the setup with the QPC formed in BLG subject to the in-plane magnetic field $B$. In experiment, the magnetic field $B$ is oriented in plane with an angle of approximately $45^{\circ}$ with regard to the black dashed line. (b) Cross-section of the setup along the dashed line in left panel. The QPC is tunable by the split gates (SG), back gate (BG), and top gate (TG).  }
	\label{fig:setup}
\end{figure*}

One rather notorious phenomenon, where interaction effects show up in transport measurements, is the appearance of an additional shoulder in the quantized conductance of QPCs below the lowest plateau. It is commonly known as the $0.7$ conductance anomaly, since, in systems with spin degeneracy, it is usually observed close to the value of the conductance
$$ G\approx 0.7\times 2\, \frac{e^2}{h}.$$ 
This feature was first recognized in a GaAs QPC in Ref.~\cite{Thomas1996} and, while it is fair to say that there is still no commonly accepted theory explaining all aspects of the anomaly, there are several microscopic theories that are capable of capturing some salient features of the phenomenon. These theories invoke various distinct physical mechanisms driven by electron-electron correlations, such as variants of the Kondo effect \cite{Meir2002, Cronenwett2002, Golub2006, Koop2007, Tripathi2008, Brun2014}, Wigner crystallization \cite{Flambaum2000, Sushkov2001, Liu2009}, and other interaction-based mechanisms \cite{Sushkov2003, Sloggett2008, Lunde2009, Tkachenko2013, Bauer2013, Heyder2015, Iagallo2015}. 
In particular, there are studies investigating the influence of the QPC barrier on electron-electron interaction effects perturbatively. On a very simplistic level, considering local interaction, only the Hartree type processes involving electrons with opposite spin contribute, leading to an effective blocking of the channel for one spin species for a certain amount of time and thus lowered conductance. Since interaction effects are enhanced at low densities, such type of effects would be strongest in the lowest quantization subband.

In this work, we study the conductance of a BLG QPC for in-plane magnetic field orientation. We start with presenting our experimental results (Sec.~\ref{Sec:2}), which were obtained in the same sample as in Ref.~ \cite{Kraft2018b}, but in another cool down for changing the sample orientation within the magnet.
In particular, we demonstrate the importance of interaction effects in the lowest size-quantized subbands by measuring the renormalized Land\'{e} $g$-factor governing the Zeeman splitting of the subbands. This motivates us to employ the picture based on the interaction-induced spontaneous polarization of spin or valley degrees of freedom to describe the shoulder-like features in the conductance. After a short reminder on the band structure of BLG and, especially, the influence of external gating on the gap and the densities (Sec.~\ref{Sec:3}), we discuss the conductance of the BLG QPC. In Sec.~\ref{Sec:4} we detail an extension of a phenomenological model for the $0.7$ anomaly proposed in Ref.~\cite{Bruus2001} to BLG. Within this framework, we investigate all possible scenarios in order to find the one most likely to be present in this experiment. We do not explicitly consider any microscopic model of the anomaly, but, instead, assume that some sort of interaction-induced spin and/or valley splitting is present at zero magnetic field and investigate the consequences of possible types of splitting on the conductance in increasing magnetic field. In fact, the assumed polarization does not need to be static, it just needs to fluctuate slowly compared to the traveling time through the constriction, which according to Ref.~\cite{Schimmel2017} is indeed fulfilled. By comparing our experimental results with these scenarios (Sec.~\ref{Sec:5}), we conclude that our sample shows spontaneous spin polarization but no valley splitting. Our findings are summarized in Sec.~\ref{Sec:6}, and technical details are described in Appendices.

\section{Experimental results}
\label{Sec:2}
\subsection{Fabrication and characterization}
For this experiment, we have used the same BLG device as presented in  Ref.~\cite{Kraft2018b},
see Fig.~\ref{fig:setup}. The chosen gate configurations is $V_\text{BG}=10$ V (back-gate voltage) and $V_\text{SG}=-12$ V (split-gate voltage). This setup differs from the one used in Ref.~\cite{Kraft2018a} for the study of the supercurrent confinement in BLG QPC by an addition of an overall top gate. The device consists of a hBN-BLG-hBN heterostructure, which is edge contacted with $\mathrm{Ti}$/$\mathrm{Al}$ electrodes. The thickness of the top and bottom hBN layers of the sandwich are $38$~nm and $35$~nm, respectively. The sandwich is placed onto a pre-patterned back gate, which is designed on a sapphire substrate that is, in turn, covered by an additional layer of the dielectric $\mathrm{Al_2O_3}$. 
The magnetic field was applied in the plane of the BLG layer. The measurements were performed under the same experimental condition as in Ref.~\cite{Kraft2018b}, but in a different cool down, with the magnetic field oriented in the plane of the BLG (at approximately 45° from the current direction).

The QPC in BLG is engineered electrostatically by means of the split gate placed on top of the device and the whole sample is covered in an extra layer of $\mathrm{Al_2O_3}$ with $30$ nm thickness before adding the overall top gate made from $\mathrm{Ti}$/$\mathrm{Cu}$. The measurements were performed at either $20$mK or $4$K in a $\mathrm{{}^3He/{}^4He}$ dilution refrigerator BF-LD250 from BlueFors. A two-terminal configuration was used employing the standard low-frequency ($\approx 13$Hz) lock-in technique, with an AC-excitation ranging from $1$ to $20\mu$V.
For further details of the characterization of the sample the reader is referred to the Supplemental Material in Ref.~\cite{Kraft2018b}. Figure 8 of the Supplemental Material in Ref.~\cite{Kraft2018b} also shows the finite-bias measurements used to extract the gate-coupling parameter.

To the best of our knowledge, there are two papers by other groups that have investigated similar setups, namely Ref.~\cite{Lee2020} and Ref.~\cite{Banszerus2020}. While both these papers also studied transport through a BLG QPC, the confinement conditions there were different from those in our setup. This difference might be crucial for observation interaction effects, including the 0.7 anomaly. Specifically, in the present work, the QPC is formed by split gates of a physical width $w\approx$ 65~nm. Because of the additional layers of $\mathrm{Al_2O_3}$, the distances between the channel and the global back and top gate are 55~nm and 68~nm, respectively. In Ref.~\cite{Lee2020}, the physical width of the split gates is 120~nm, while the distance to the back gate and split gate is not specified. Since Ref.~\cite{Lee2020} did specify that the BLG is encapsulated in hBN, the distance to the back gate and the split gate is likely of the order of 30~nm, with an additional 35~nm of $\mathrm{Al_2O_3}$ between split gates and local top gate. Similarly, Ref. \cite{Banszerus2020} stated a width of 250~nm, a distance of 25~nm to the back gate (and, probably, a similar one to the split gates), and additionally 25~nm of $\mathrm{Al_2O_3}$ between split gates and a local top gate. This means, that our channel is a lot narrower, confinement a lot stronger, and, thus, the density of states way larger, which enhances all interaction effects.

 Moreover, interaction effects in Refs.~\cite{Lee2020,Banszerus2020} should be more strongly suppressed by the top and bottom gate, which are closer than the typical distance of interacting electrons within the constriction. It is worth noting that Ref.~\cite{Kim2020} stated that gates need to be closer than a few nanometer, to fully suppress electron-electron interaction in graphene and BLG. At this point, it should be mentioned that, depending on the exact shape of the constriction, the 0.7 shoulder can appear at different conductance values (for example, at 0.5 $e^2/h$ \cite{Bauer2013,Heyder2015}), which would fit with the alleged spin-orbit gap of Ref.~\cite{Banszerus2020}. 

 The global back gate that covers also parts of the leads in our device leads to a smoother coupling in the QPC region, while also modifying the band structure and gap in the non-QPC regions. As has been shown, for example, in Refs.~\cite{Bauer2013,Heyder2015}, both the presence and shape of the 0.7 anomaly depend rather strongly on the exact constriction profile, so that a smoother constriction region might be necessary for its appearance. This also applies to the larger parameter space we explore by varying our split gate and back gate not only along the direction of zero displacement field.  Lastly, we want to point out that most of our reported results are based on the three lowest size quantized levels, which are not even resolved in Ref.~\cite{Lee2020}, while Ref.~\cite{Banszerus2020} does not reach full pinch-off.

 \begin{figure*}
\centering
	\includegraphics[width=
	\textwidth]{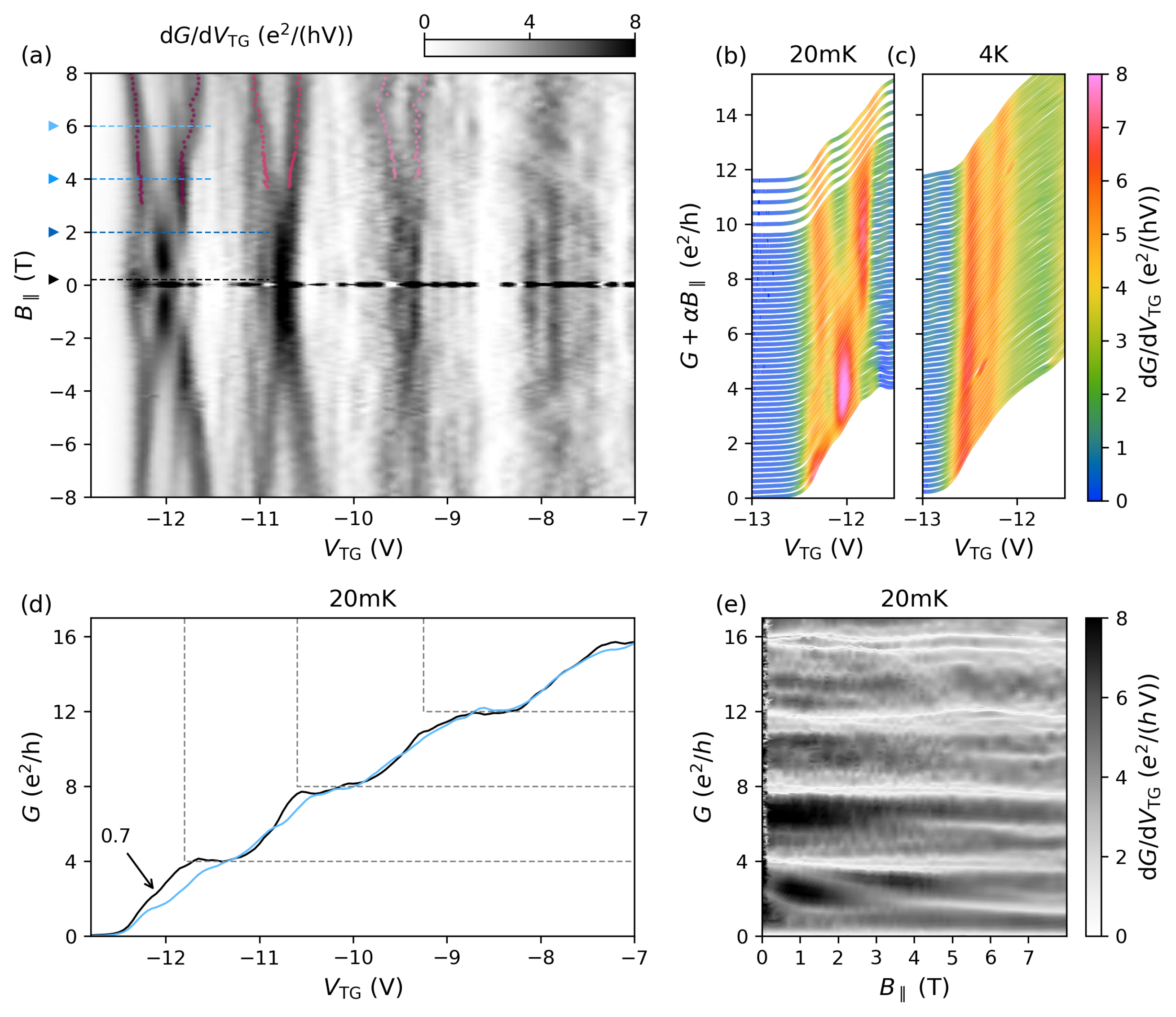}
	\caption{Measured conductance of the QPC in BLG for $V_\text{BG}=10$~V and $V_\text{SG}=-12$~V. (a): Differentiated differential conductance as a function of the top-gate voltage $V_\mathrm{TG}$ and in-plane magnetic field $B$ for temperature $20$\,mK. Plateaus of the conductance corresponds to bright regions, steps correspond to dark region. The map scale is cut at $0\,e^2/hV$ and $8\,e^2/hV$ to bring out the details. The black dashed line corresponds to $B=0.2$\,T, the lines of blue tones to $2,4,6$\,T. The dots of different shades of pink mark the development of the spin subbands used to extract the Zeeman splitting and the effective  Land\'{e} $g$-factors. (b) and (c): Cubic spline fit of the differential conductance $G$ as a function of the $V_\mathrm{TG}$ in elevating $B$ for $20$\,mK and $4$\,K, respectively. The curves are shifted vertically with $\alpha = 2{e^2}/{h\mathrm{T}}$ and colored according to their first derivative. (d): Differential conductance $G$ as a function of $V_\mathrm{TG}$ at $20$\,mK for $B=0.2$\,T (black curve) and $B=6$\,T (light blue). The arrow marks the additional shoulder, which we identify as a 0.7 anomaly. (e): Differentiated differential conductance as a function of $B$ and conductance $G$ for $20$\,mK. Plateaus of the conductance correspond to bright regions, slopes correspond to dark regions.}
	\label{fig:exp}
\end{figure*}

\subsection{Conductance}

We start by investigating the dependence of the conductance on the magnetic field and top-gate voltage. 
Figures~\ref{fig:exp}(a)-(d) show the experimental data at temperature $20$ mK. In Fig.~\ref{fig:exp}(d), the conductance is shown as a function of the top-gate voltage $V_\mathrm{TG}$ for two different values of in-plane magnetic fields $B_\parallel$. The black curve corresponds to $B_\parallel=0.2$ T and the light-blue one to $B_\parallel=6$ T as marked in Fig.~\ref{fig:exp}(a). The light-blue curve highlights the appearance of additional half-step conductance plateaus in high in-plane magnetic fields. The black curves contains a shoulder marked by the arrow, which we will attribute to the 0.7 conductance anomaly. We note that the valley degeneracy is apparently not affected by the application of the in-plane magnetic field, and the Zeeman spin-split subbands remain degenerate in the two valleys $K$ and $K'$. Since the aluminum leads are superconducting at $20$\,mK, a finite magnetic field is needed to kill this effect and curves below $0.2$\,T show influence of the superconducting leads, cf. Appendix \ref{sec:addPlots}.

Cubic spline fits of the conductance for all measured values of magnetic field between $0.2$\,T and $6$\,T are shown in Fig.~\ref{fig:exp}(b) and Fig.~\ref{fig:exp}(c) for temperatures $20$\,mK and $4$\,K, respectively. Curves in both figures are shifted vertically for clarity and colored according to their first derivative. For both temperatures, there are two regions of steep incline (orange-red) for high magnetic field, corresponding to the chemical potential crossing through the spin-split bands. The splitting is both sharper and higher for the lower temperature, and plateaus are flatter there as well. The lower spin-subband stays roughly at the same value of $V_\mathrm{TG}$.

\begin{figure*}
\centering
	\includegraphics[width=
	\textwidth]{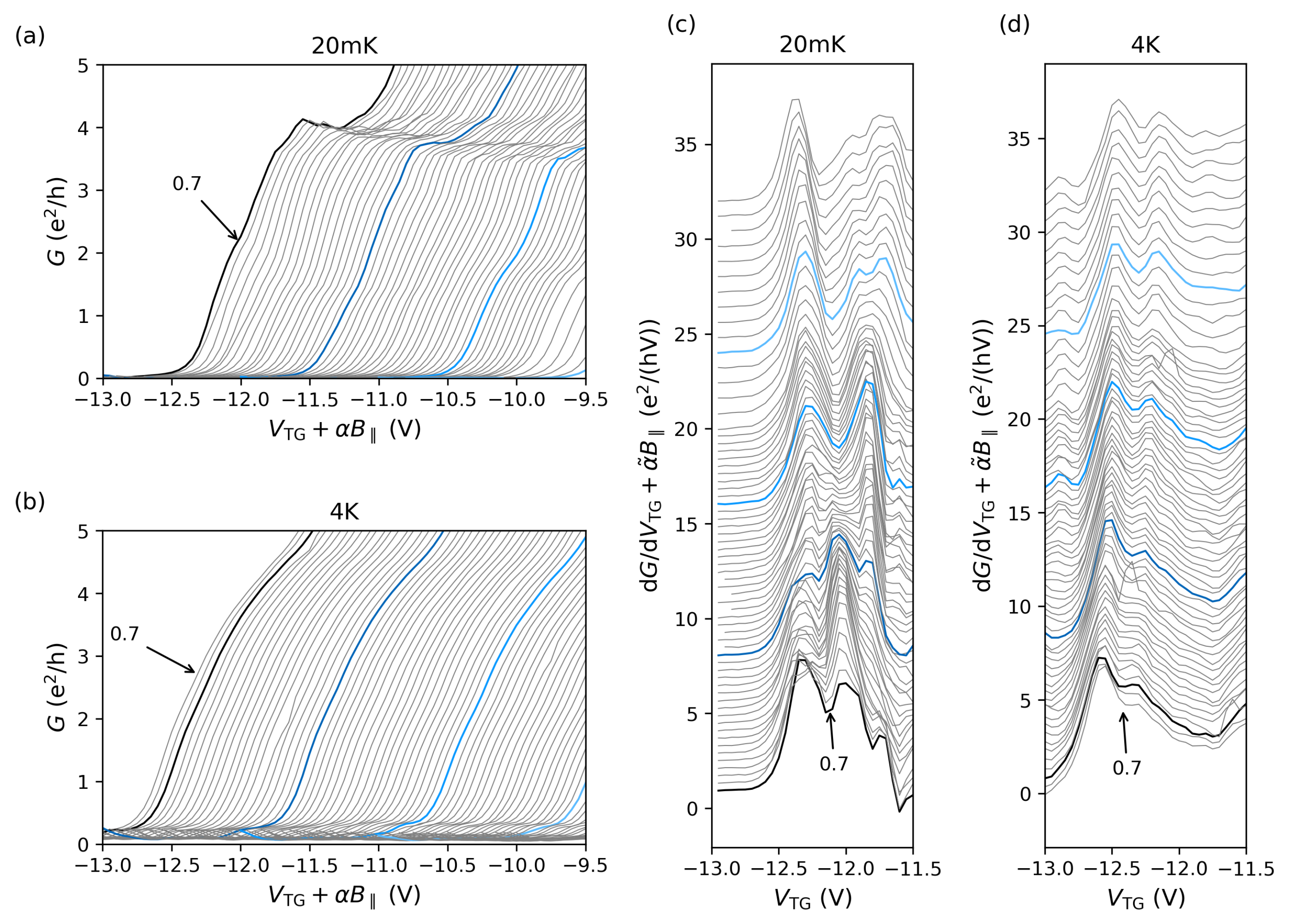}
	\caption{(a) and (b): Zoom-in view for the lowest plateau of the differential conductance $G$ at $20$\,mK and $4$\,K, respectively, as a function of $V_\mathrm{TG}$ with horizontal shifts for different values of magnetic field, $\alpha = 0.5$ V/T. (c) and (d): Differentiated differential conductance as a function of $V_\mathrm{TG}$ for $20$\,mK and $4$\,K respectively. Curves for different magnetic field values are shifted vertically with $\tilde{\alpha}=4\mathrm{e^2/(hV T)}$. In all panels, the colored curves correspond to the magnetic field values introduced in Fig.~\ref{fig:exp}(a).}
	\label{fig:zoom}
\end{figure*}

Figure~\ref{fig:exp}(a) shows a grayscale map of the differentiated differential conductance $dG/dV_\mathrm{TG}$ as a function of top gate voltage $V_\mathrm{TG}$ and in-plane magnetic field $B_{\parallel}$ for $T=20$\,mK. Transitions across 1D subband edges appear as dark lines, while conductance plateaus are visible as light regions in between. One clearly sees the four well-resolved conductance plateaus. These are separated by the three regions corresponding to the 1D subbands, which are split roughly symmetrically with the applied in-plane field for higher bands.  This corresponds to the evolution from the spin-degenerate into spin-split energy levels. The lifting of the spin degeneracy occurs for the lowest three subbands, where the confinement and interactions are the strongest. 

Figure ~\ref{fig:exp}(e) shows the same data as ~\ref{fig:exp}(a), but as a function of $B_\parallel$ and $G$. The bright horizontal lines at multiples of $4 e^2/h$  correspond to the spin- and valley-degenerate conductance quantization plateaus for zero magnetic field, the additional half-integer multiples for higher magnetic fields correspond to the spin-split plateaus due to the Zeeman effect.

\subsection{Extra features of the conductance}

Additionally, we note the presence of a shoulder-like feature below the lowest conductance plateau at about $G=2.5\,e^2/h$, similar to the $0.7$ structure described in many other materials \cite{Micolich2011}, which develops into the lowest spin-split subband at $G=2\,e^2/h$. This feature is well visible in the black curves in Figs. \ref{fig:exp}(a) and \ref{fig:exp}(d). Since flatter parts of the conductance correspond to brighter color in Figs.~\ref{fig:exp}(c) and \ref{fig:exp}(f), it corresponds to a bright region in between the zeroth and first plateau, i.e., within the darker region to the left of $V_\text{TG}=-12$\,V, making it look like a spin splitting of the 1D subbands at zero magnetic field.

This additional feature is also visible in Fig.~\ref{fig:zoom}, which shows cadence plots of the conductance at $20$\,mK and $4$\,K in Fig.~\ref{fig:zoom}(a) and (b), respectively, and of the derivative of the conductance at $20$\,mK and $4$\,K in Fig.~\ref{fig:zoom}(c) and (d), respectively. In all cases, only the lowest band is shown. The cadence plots for a larger range of conductance variation are shown in Appendix~\ref{sec:addPlots}. The colored curves correspond to the values of magnetic field marked in Figure ~\ref{fig:exp}(a). In the black curves in both  Figure ~\ref{fig:zoom}(a) and (b) the is an additional shoulder at around $2.5 e^2/h$, which develops into the spin split plateau for higher magnetic fields. In the cadence plots of the conductance Figure ~\ref{fig:zoom}(c) and (d) this shoulder corresponds to an additional peak. which clearly develops into the spin split peak for $4$ K whereas this transition is somewhat obscured by yet another feature at $20$ mK. We identify this obscuring feature as part of a larger oscillation pattern discussed later. Similar plots are shown in Ref.~\cite{Koop2007} for GaAs, where the observed behavior was attributed to the $0.7$ structure. 

The extra feature cannot be an effect caused by the finite magnetic field needed to kill superconductivity, since it is not located on the imaginary line extending the Zeeman splitting down to small magnetic fields. Instead, a finite magnetic field is needed to bring this feature down to the spin-split value. Moreover, this feature is seen already at zero magnetic field in Figs.~\ref{fig:exp}(c) and \ref{fig:zoom}(b) and (d) at higher temperature, where the contacts are not superconducting. At stronger magnetic fields, $B_\parallel\gtrsim 4\,$T, this feature merges with the shoulder that, at the lowest magnetic fields, splits off the lowest main conductance quantization plateau at $G=4\,e^2/h$ and goes down to form a plateau slightly below $G = 2\,e^2/h$.
This behavior is clearly observed as the evolution of the red region above $V_\mathrm{TG}\approx -12$\,V in Fig.~\ref{fig:exp}(b). The merging of the two shoulders is also evident in
Fig.~\ref{fig:exp}(a) as an intersection of the two bright regions at $B_\parallel\approx 4\,$T
and $V_\mathrm{TG}\approx -12$\,V. 

Finally, there are additional oscillations in the conductance (of which the obscuring feature in Fig.~\ref{fig:zoom}(c) is one), which are most visible close to conductance plateaus in Figs.~\ref{fig:exp} (d).  These appear as vertical lines in Fig.~\ref{fig:exp}(a) and are less visible for the higher temperature in Fig.~\ref{fig:zoom}(b). Most notably, a maximum of such an oscillation is seen to go straight through one of the spin-split bands of the lowest 1D subband in Fig.~\ref{fig:exp}(a) and (b) and Fig.~\ref{fig:zoom}(c), starting at around $-12$V and $0$T in the lowest plateau, crossing one spin subband at around $3$T, and ending up in the $0.5\,e^2/h$ plateau for higher magnetic field. Similar oscillations appear at other voltages in a regular fashion.

\subsection{Effective Land\'{e} \textit{g}-factor}

\begin{figure}
	\centering
	\includegraphics[width=\columnwidth]{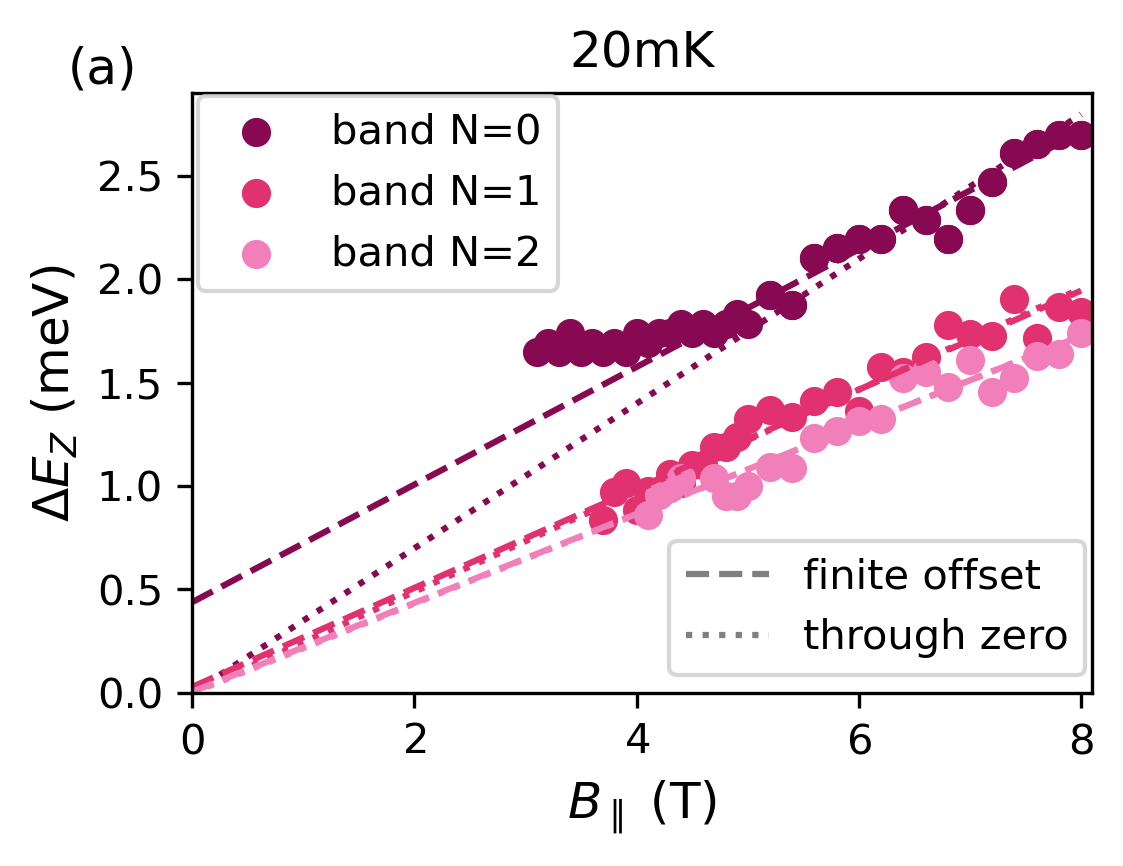}
		\includegraphics[width=\columnwidth]{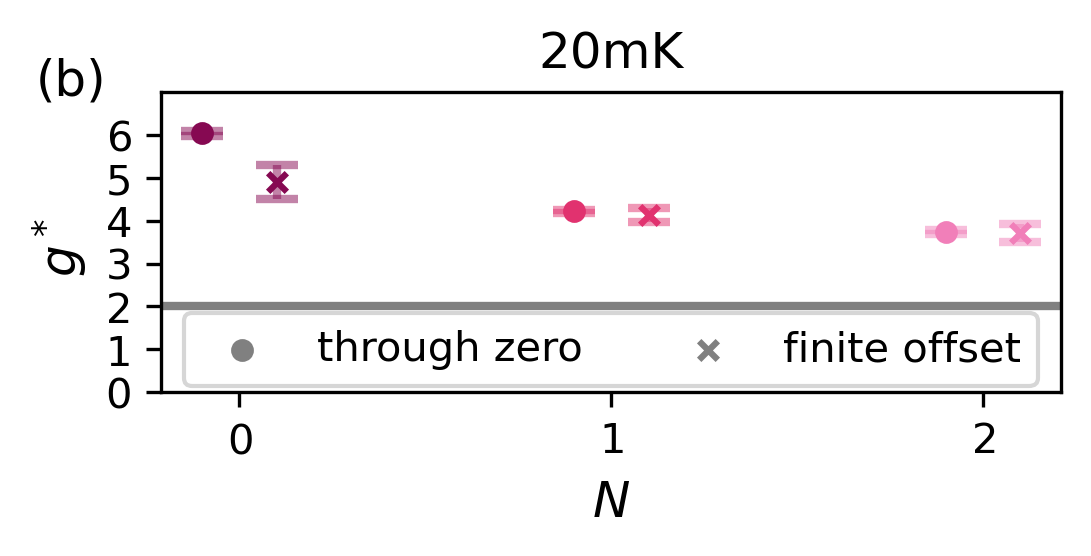}
	\caption{(a) Zeeman energy splitting $\Delta E_Z$ as a function of magnetic field $B$ for various 1D subbands. Dashed lines show best linear fits of high-field data points, dotted ones are linear fits going through zero at zero magnetic field. (b) Extracted effective Land\'{e} $g$-factors $|g^*|$ for the three quantization subbands [solid dots of the colors corresponding to legend in panel (a)], obtained for the in-plane magnetic field. The gray line indicates the value bare 2D $g$-factor $g=2$ for BLG. The error bars mark the 1$\sigma$ intervals from the two performed fits shown in panel (a). The dotted points correspond to the dotted lines above, the crosses to the dashed lines. All parameters are given in Table~\ref{tab:parameters}. 
	}
	\label{fig:3}
\end{figure}

From the spin splitting of the 1D subbands marked in pink in Fig.~\ref{fig:exp}(a) we extract the Zeeman energy splitting $\Delta E_Z$ by converting the top-gate voltage $V_\mathrm{TG}$ into energy, using the splitting rate of the energy levels in source-drain bias measurements~\cite{Kraft2018b}, as described in Ref.~\cite{Patel1990, Patel1991, MartinMoreno1992, Danneau2006}. The confinement in this cooldown, $V_\text{BG}=10$ V and $V_\text{SG}=-12$ V, does not exactly correspond to the setup in the source-drain measurement, where $V_\text{SG}=-11.6$~V.
We observed a good agreement between the two measurements in Ref.~\cite{Kraft2018b}, which had a bigger difference in the confining potentials. Most importantly, the extracted gate coupling is the same for all nine visible subbands. Thus, we expect this value to be a very good fit here as well and use $$E=\alpha_\text{TG}\, e \left(V_\text{TG}-V_\text{TG}^{(0)}\right),
\quad \alpha_\text{TG}=3.8\times 10^{-3}.$$ 
The obtained value of $\Delta E_Z$ for each of the three lowest subbabnds is plotted in Fig.~\ref{fig:3}(a) as a function of magnetic field, revealing linearly increasing Zeeman energy splittings. Remarkably, in case of the $N=0$ subband, the Zeeman splitting shows a linear behavior only for $B_\parallel\gtrsim 5$~T, whereas at smaller fields an almost constant splitting is observed.

This saturation effect can be linked to the observed additional shoulder in the conductance curves in Figs.~\ref{fig:exp}(b) and \ref{fig:exp}(c) at not too strong magnetic fields. The plateau in the Zeeman splitting corresponds to the magnetic fields below $4$\,T in Fig.~\ref{fig:exp}(a), where the bright region to the left of $V_\text{TG}=-12$\,V disappears.
One can either fit the dependence of $\Delta E_Z$ on the magnetic field requiring a vanishing splitting extrapolated to zero $B$ or not (using then the best linear fit at high magnetic field). 
In the latter case, a finite intercept at $\Delta E_Z\approx 1.7$\,meV is observed for $N=0$ subband at $B_\parallel=0$, unlike the cases $N=1$ and $N=2$, which extrapolate to close to zero energy splitting.
This suggests that a spontaneous spin splitting occurs for the $N=0$ subband, where the effects of the interaction and confinement are expected to be the most prominent. 
Fitting with a finite intercept, as was done, e.g., in Ref.~\cite{Koop2007}, establishes a bound on zero-field splitting without interaction effects. One should note that this splitting is fully obscured by the much larger, interaction-induced 0.7 anomaly that produces a much larger value of the zero-$B$ splitting.

\begin{table}[ht!]
    \begin{ruledtabular}
    \begin{tabular}{llll}
    subband     & 0 & 1 & 2  \\
         \hline
      $g^*$ (no offset) &6.04(6) & 4.22(4) & 3.73(4)\\
       $g^*$ (finite offset) &4.91(40) & 4.14(17) & 3.72(22)\\
     offset in meV & 0.438(154) & 0.0288(546) & 0.00506(755)
    \end{tabular}
    \caption{Parameters extracted from the fit of the splitting in magnetic field. The uncertainty corresponds to the  uncertainty of the fit, i.e., a $1\sigma$ interval.}
    \label{tab:parameters}
    \end{ruledtabular}
\end{table}

From the slopes of the Zeeman splitting in Fig.~\ref{fig:3}(a), we find the (independent of the magnetic field) values of effective Land\'{e} $g$-factors $|g^*_N|$ for each of the subbands, shown in Fig.~\ref{fig:3}(b), see Table~\ref{tab:parameters}. These values are obtained by taking the linear fit to the splitting with and without a finite intercept. For both fits we use the 1$\sigma$ intervals to obtain error bars.  The obtained $|g^*_N|$ values are increasingly enhanced for lower subbands compared to the bare 2D $g$-factor $g=2$, with a maximum enhancement by a factor of about 2--3 for $N=0$. This observation also supports the idea of an enhanced role of interaction effects for the $N=0$ subband. Independent of the exact value of the gate coupling, this enhancement only relies on a gate coupling that is the same for the three subbands. This enhancement is seen both for a fit with finite intercept or without. Since the reported Kane-Mele spin-orbit gap of $0.04-0.08$~meV is in between the finite and the vanishing intercept, it would also not change the resulting enhancement of the Land\'{e} $g$-factor by more than a few percent.

\section{Theoretical model}
\label{Sec:3}
The quantization of conductance in a QPC is a well-known experimental proof of the possibility of confining charge carriers and it clearly shows their quantum nature \cite{Wees1988a}. What makes BLG an interesting platform for such measurements, is its additional valley degree of freedom and the high electrostatic tunability of its band gap \cite{Oostinga2008,Allen2012}. In this section, we discuss the effects of the applied gate voltages on the band structure and, thus, on the observed conductance within the essentially non-interacting model (interaction here is taken into account only through the self-consistent screening of the gate potentials).

\subsection{Effective Hamiltonian and dispersion of BLG}

We describe the low-energy properties of BLG relevant for the transport measurements in the QPC geometry by the effective two-band Hamiltonian, see Ref.~\cite{McCann2006a}. The details of this approximation are given in Appendix~\ref{sec:eff2DModel}. The two-band matrix Hamiltonian, acting in the space of the pseudospin degree of freedom (Pauli matrices $\hat{\sigma}$) combined with the Zeeman interaction in the spin space (Pauli matrices $\hat{s}$), has the form
\begin{align}
\hat{H}&=\left(\hat{H}_0+\hat{H}_M \right)\otimes \hat{s}_0+\frac{1}{2}\Delta E_Z\hat{\sigma}_0\otimes\hat{s}_z \label{eq:BLGHam},
\\
\hat{H}_0&=-\frac{1}{2m}\begin{pmatrix}
0 & (\pi^{\dagger})^{2}\\
\pi^{2} & 0
\end{pmatrix}-\frac{U}{2} \begin{pmatrix}
1 & 0\\
0 & -1
\end{pmatrix}
\label{eq:Ham0},
\\
\hat{H}_M&=\frac{U v^2}{\gamma_1^2}\begin{pmatrix}
\pi^{\dagger}\pi & 0\\
0 & -\pi\pi^{\dagger}
\end{pmatrix}
\label{eq:MexicanHam}.
\end{align}
Here, $\pi=\xi p_x+ip_y$ is the kinetic momentum,
with ${\xi=\pm}$ referring to the $K_\pm$ valley.
Here, we disregard possible spin-orbit coupling, which is a small effect at the energy scales of the experiment and not capable of explaining the zero-field splitting or the magnetic field behavior we observe, as seen by the obtained zero-field splitting of Fig.~\ref{fig:3}(a). We will return to this issue again below. 

In what follows, we disregard the Mexican-hat term $\hat{H}_M$ that develops for finite layer asymmetry $U$ as discussed in Ref.~\cite{McCann2013}. We also neglect the skew interlayer hopping, which leads to trigonal warping~\cite{McCann2013,Knothe2018}. The effect of these subtle features of the BLG spectrum on the conductance of a QPC in in-plane magnetic field will be discussed elsewhere. Here, we adopt the simplest model that, as we demonstrate below, is capable of describing the salient features of the conductance. 

Clearly, we have to distinguish the two spatial regions in our physical sample. Away from the split gates there is no confinement and electrons feel an approximately constant top-gate and back-gate voltage. Close to the split gate, the shape of the confinement leads to a non-trivial, spatially dependent effective top-gate voltage.

The dispersion of the spin $\sigma=\uparrow,\downarrow$ band for the low-energy 
Hamiltonian (\ref{eq:BLGHam}) without the 
Mexican-hat feature~(\ref{eq:MexicanHam}) 
is given by 
\begin{align}
E_\sigma = 
\pm\sqrt{\frac{U^2}{4}+\frac{\hbar^4 k^4}{4m^2}}
+\frac{\sigma}{2}\Delta E_Z. \label{eq:2Ddispersion}
\end{align}
This corresponds to a 2D density for spin projection $\sigma$:
\begin{align}
n^\mathrm{2D}_\sigma(\mu)
&= 2\frac{m}{8\pi \hbar^2}\sqrt{ 4\left(\mu-\frac{\sigma}{2}\Delta E_Z\right)^2-U^2},
\label{eq:2Ddensity}
\end{align}
where the factor of $2$ accounts for the valley degree of freedom and the chemical potential $\mu$ is measured with respect to the middle of the asymmetry gap. For a small Zeeman splitting,
$\Delta E_Z\ll \sqrt{4\mu^2-U^2}$, one can use
the expansion
\begin{align}
n^\mathrm{2D}_\sigma(\mu)
&\approx \frac{m}{4\pi \hbar^2}\left(\sqrt{ 4\mu^2-U^2}-\frac{2\sigma \mu\Delta E_Z }{\sqrt{4\mu^2-U^2}}\right). 
\label{eq:2Ddensity-expanded}
\end{align}
This expansion tells us that the effect of the Zeeman splitting on the density 
is enhanced when the chemical potential is close to the gap.
The total density $n^{2D}=\sum_{\sigma=\pm} n^\text{2D}_\sigma $ is, to first order in $\Delta E_Z$, independent of magnetic field, and we get for the chemical potential in weak fields:
\begin{align}
    \mu(n^\text{2D})&=\frac{\sqrt{m^2 U^2+4 \pi ^2 (n^\text{2D})^2 \hbar ^4}}{2 m}\nonumber\\
   & -\frac{ m^3 U^2 (\Delta E_Z)^2}{16 \pi ^2 (n^\text{2D})^2 \hbar ^4 \sqrt{m^2 U^2+4 \pi ^2 (n^\text{2D})^2 \hbar ^4}}.
    \label{eq:chemPot}
\end{align}

\subsection{Controlling BLG with gates \label{Sec:IIIB}}

In the 2D regions away from the QPC, the effect of a constant back-gate and top-gate voltage is described by the self-consistent gap equation~\cite{McCann2006, McCann2013}. 
The total density $n=n_\uparrow+n_\downarrow$ is electrostatically determined by the gates and  given by
\begin{align}
n = \frac{\varepsilon_0\varepsilon_\text{BG} V_\mathrm{BG}}{e L_\text{BG}}+\frac{\varepsilon_0\varepsilon_\text{TG} V_\mathrm{TG}}{e L_\text{TG}} \label{eq:electrodensity}.
\end{align}
Here, $\varepsilon_0$ is the vacuum permittivity,  $L_\text{BG}$ ($L_\text{TG}$) is the distance from the BLG plane to the back gate (top gate), and $\varepsilon_\text{BG}$, $\varepsilon_\text{TG}$ are the relative dielectric constants of the material between BLG and the back gate and top gate, respectively. 
In the absence of screening, the interlayer asymmetry factor $U$ is given by 
\begin{align}
U_\mathrm{ext} = \frac{e c_0}{2\varepsilon_r}\left(\frac{\varepsilon_\text{BG}}{L_\text{BG}}V_\text{BG}-\frac{\varepsilon_\text{TG}}{L_\text{TG}}V_\text{TG}\right),
\label{Uext}
\end{align}
where $c_0$ is the distance between the two BLG planes and $\varepsilon_r$ is the relative permittivity between these sheets.

\begin{figure*}[htp]
	\centering
	\includegraphics[width=0.95
	\textwidth]{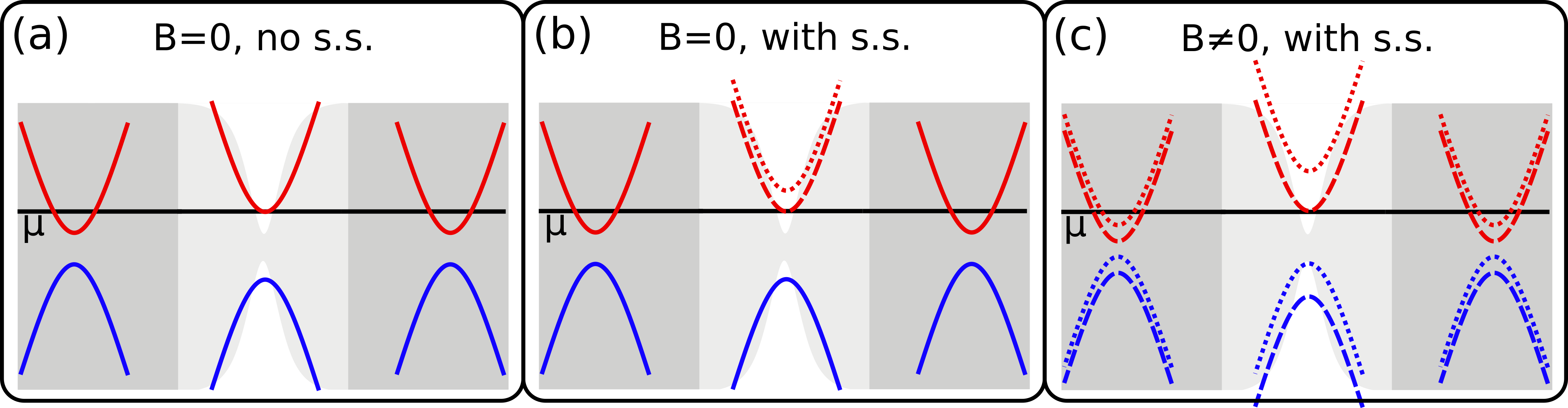}
	\caption{(a) and (b): Band structure in the 2D regions (dark gray) and inside the constriction (light gray) for zero magnetic field, with and without spontaneous splitting (s.s.), respectively. Here, the spontaneous splitting is between spin subbands shown as dashed or dotted lines. Since it requires the proximity of the chemical potential, there is no s.s. in the valence band. Because of the split-gate stray fields, the asymmetry gap inside the constriction is larger. Energies are measured with respect to the middle of the gap in the 2D region. (c): The magnetic field introduces a spin splitting of the bands with respect to the $B=0$ band bottom, which leaves the middle of the gap at the same value. The splitting inside the constriction is larger, since the Land\'e $g$-factor is enhanced. As observed experimentally, the Zeeman splitting inside the constriction is not symmetric. To the lowest order in  a weak magnetic field, the chemical potential for a fixed total 2D density is independent of magnetic field, see Eq.~(\ref{eq:2Ddensity-expanded}).}
	\label{fig:constriction}
\end{figure*}

Since the two layers of BLG screen the effect of the closer gate for the other BLG plane depending on their density and thus the felt voltage, the actual asymmetry as a function of the density is given by the self-consistent equation \cite{McCann2013}
\begin{align}
U(n)&=\dfrac{U_\mathrm{ext}}{1-\frac{\Lambda}{2}
\ln \left(\frac{|n|}{2 n_\perp}+\frac{1}{2}\sqrt{\frac{n^2}{n^2_\perp}
+\frac{U^2}{4\gamma_1^2}}
\right)}
\nonumber
\\
&\approx U_\mathrm{ext}\left[1-\frac{\Lambda}{2}\ln \left(\frac{|n|}{ n_\perp}\right)\right]^{-1}\label{eq:asymmetry},
\\
n_\perp &=\frac{\gamma_1^2}{\pi\hbar^2 v^2} ,
\qquad \Lambda = \frac{c_0e^2n_\perp}{2\gamma_1\varepsilon_0\varepsilon_r}.
\end{align}
Thus, changing the top-gate voltage tunes the density $n$ according to Eq.~(\ref{eq:electrodensity}),  which, in turn, influences the asymmetry factor $U$ according to Eq.~(\ref{eq:asymmetry}) and hence the dispersion (\ref{eq:2Ddispersion}) and the chemical potential according to Eq.~(\ref{eq:chemPot}). This chemical potential remains constant over the whole sample, including the QPC constriction, where the density is no longer given by Eq.~(\ref{eq:2Ddensity}):
\begin{align}
    \mu(V_\text{BG},V_\text{TG})&\simeq 
    \sqrt{
    \left(
    \frac{U_\text{ext}}
    {2-\Lambda\ln \frac{|n^\text{2D}|}{n_\perp}
    }
    \right)^2+ \left(\frac{\pi\hbar^2n^\text{2D}}{m}\right)^2}.
\end{align}
Here, the chemical potential depends on $V_\text{BG}$ and $V_\text{TG}$ through the corresponding dependence of the 2D density, Eq.~(\ref{eq:electrodensity}), and the dependence of $U_\text{ext}$, Eq.~(\ref{Uext}).

In the experiment, the combination of back-gate and split-gate voltages is used to open a gap $U$ under the constricted region and tune the chemical potential inside this gap, as shown in Ref. \cite{Kraft2018b}, and thus form the QPC, see Fig.~\ref{fig:constriction}. The overall top gate is used to tune into the low-density regime, where the observation of conductance quantization is possible~ \cite{Kraft2018b}. Importantly, for fixed back-gate and split-gate voltages, like in the experimental setup, the top-gate voltage tunes the electronic density in the sample linearly \cite{McCann2013}.

As proposed in Ref.~\cite{Buettiker1990}, we  model the QPC by projecting the 2D problem onto a one-dimensional one. The procedure for a standard Schr{\"o}dinger equation is described in Appendix \ref{sec:projSch}. A generalization of the method to BLG is discussed in Appendix~\ref{sec:projBLG}. The quantization of conductance is already visible in the simplest approximation of hard-wall boundary conditions, as we will show now. In the case of a channel of width $W$, the dispersion relation for the longitudinal wavevector $k$ resulting from Eq.~(\ref{eq:BLGHam}) takes the form \cite{Kraft2018b}
\begin{align}
E_{N,\sigma}(k)&=\pm \sqrt{\frac{U^2}{4}+\frac{\hbar^4}{4m^2}q_N^4(k)}+\frac{\sigma}{2}\Delta E_Z, \label{eq:BLGQuantDisp}\\
q_N^2(k)&=k^2+\left(\frac{N\pi}{W}\right)^2,
\end{align}
where $N=0,1,2\ldots$ labels the size-quantized bands.
While the case $N=0$, strictly speaking, requires a different choice of boundary condition, we still chose to investigate the effect of the resulting $k^4$ dispersion, which one would also get in the 2D setup. It will turn out, that the choice of any non-linear dispersion does not have qualitative consequences for the 0.7 effect. 
Note that $U$ in Eq.~(\ref{eq:BLGQuantDisp}) differs from the 2D expression (\ref{eq:asymmetry}), since the screening in a 1D channel differs from that in the unconfined regions of BLG. We also note that the channel width is affected in a non-trivial way by $V_\text{BG}$ and $V_\text{TG}$.

The lowest band is, to leading order, quartic in the momentum, so that the zero-temperature density  resulting from Eq.~(\ref{eq:BLGQuantDisp}) is given by
\begin{align}
n^\mathrm{1D}_\sigma(\mu)=\frac{2 \sqrt{2 m}}{\pi \hbar}\left[(\mu-\sigma \Delta E_Z/2)^2-U^2\right]^{1/4}, 
\label{eq:densitypermu}
\end{align}
as opposed to the square-root dependence of the 2D density~(\ref{eq:2Ddensity}). The total density in the constriction is again determined electrostatically by the gates, but the stray fields of the split gates make the evaluation of the dependence of the density on the gate voltages harder. Since the split-gate voltage is applied additionally in the constricted region, the gap there is larger and the density inside the QPC is lower than away from the barrier (Fig.~\ref{fig:constriction}), enabling the observation of the very lowest size-quantized bands.

\subsection{Conductance quantization}
We describe the conductance of the system by means of the Landauer-B{\"u}ttiker formula,
\begin{align}
G=\frac{e^2}{h}\sum_{\sigma,\xi}\int_{-\infty}^{\infty}\mathrm{d}\epsilon \mathcal{T}_{\sigma,\xi}(\epsilon)
\left[-f^{\prime}(\epsilon-\mu)\right] \label{eq:LandauerB},
\end{align}
where $\mathcal{T}_{\sigma,\xi}(\epsilon)$ is the transmission of a subband with spin $\sigma$ and valley $\xi$ and $f^{\prime}(x)$ is the derivative of the Fermi function 
$f(x)=[1+\exp(x/k_B T)]^{-1}$. 
Assuming an idealized step-function transmission coefficient, where a band contributes to $G$ as soon as it is starting to get filled, the Landauer-B{\"u}ttiker conductance is  given by
\begin{align}
G(T,B)&=2\frac{e^2}{h}\sum_{N,\sigma} f\left(E_N^{0}+\frac{1}{2}\sigma g^{*}\mu_B B-\mu\right),
\label{Eq:CondTrivial}
\end{align}
where the factor of $2$ accounts for the valley degeneracy, 
\begin{align}
E_N^{0}&=\sqrt{\frac{U^2}{4}+\left(\frac{\hbar^2}{2m}\right)^{2}\left(\frac{N\pi}{W}\right)^4}
\end{align}
is the lower band edge of band $N$ at zero magnetic field, and the Zeeman interaction is written explicitly.  

Every time the chemical potential crosses another lower band edge at finite magnetic field, the conductance makes a step of $\Delta G=2\,e^2/h$ and, for zero magnetic field, a step of $\Delta G=4\,e^2/h$. Each step has the shape of the Fermi function. The steps are separated by conductance plateaus, thus giving rise to a staircase structure seen in Fig~\ref{fig:exp} and Fig~\ref{fig:zoom}. This is the conventional conductance quantization for a QPC, with an appropriate degeneracy of the bands. In contrast to the case of an out-of-plane magnetic field~\cite{Kraft2018b,Knothe2018}, the in-plane magnetic field does not couple to the valley degree of freedom. As discussed in Ref.~\cite{VanderDonck2016}, the direct effect on the band structure is also negligible at experimentally accessible magnetic fields. Therefore, at arbitrary fields, the steps of non-interacting conductance have a factor of two corresponding to the two valleys of BLG.

\subsection{Screening and electron-electron correlations}

Electrons in the device are subject to Coulomb interaction, which is screened by the electrons themselves, by the metallic gates, and by the dielectric material.  
Let us first discuss the screening effect of the gates. 
There are three relevant length scales in the system. The first one is the physical distance between the split gate fingers is $w\approx 65$~nm and the electrostatically induced channel is smaller than that. The width of the split-gate fingers is of the order of $L\approx 300$ nm, so that we can distinguish two ranges of length scales relevant to electrostatic screening in our device.

On scales smaller or of the order of $w$, the system is truly 2D, only for larger distances it crosses over to 1D. Another relevant scale is the distance to the back gate and top gate, which are both of the order of $d\approx 55$~nm. Here, we also take into account the dielectric screening by further assuming, for simplicity, that the insulating layers in between have the same dielectric constant $\epsilon_r$ (the vacuum dielectric constant is denoted below by $\epsilon_0$). The bare, only dielectrically screened Coulomb interaction is given by its Fourier component at wave vector $\mathbf{q}$ (different in 1D and 2D cases): 
\begin{align}
    V(q)=\begin{cases}
    \dfrac{e^2}{2\epsilon_0\epsilon_r  }\dfrac{1}{q},& \text{2D},\\[0.5cm]
    \dfrac{e^2}{2\pi \epsilon_0\epsilon_r } \ln\dfrac{1}{q w},& \text{1D}.
     \end{cases}
    \end{align}

The gate-screened interaction can be found by summing up the infinite series of mirror charges. In the 2D case, this leads to
\begin{align}
    V(q)&=\frac{e^2}{2\epsilon_0\epsilon_r}
    \sum_{k=-\infty}^\infty\left(\frac{e^{-2 l|k|q}}{q}-\frac{e^{-2 |l^\prime-k l|q}}{q}\right)\\
    &=\frac{e^2}{2\epsilon_0\epsilon_r}\frac{2}{q}
    \frac{\sinh(ql-ql^\prime)\sinh(ql^\prime)}{\sinh(ql)}\\
    &=\frac{e^2}{2\epsilon_0\epsilon_r} \frac{\tanh(qd)}{q},
\end{align}
where $l=L_\text{TG}+L_\text{BG}$, $l^\prime =L_\text{BG} $ and in the last line we assumed $L_\text{TG}=L_\text{BG}=d$. This means that screening strongly alters the interaction if $qd\ll 1$. 
But in the 2D case we require $r<w$, i.e., $q>1/w$, and thus $qd\gtrsim d/w \gtrsim 1$, so that the interactions are not strongly altered by the screening of the gates.

A closer look, including the screening effects on the interaction for monolayer graphene, is discussed in Ref.~\cite{Kim2020} and reveals, that gates need to be closer than a few nanometer to really alter the interaction, which is not experimentally accessible and certainly not the case here. There, it has also been stressed that for BLG distances need to be even closer. In the 1D case the presence of the gates is relevant only on scales $x>d$ and $q<1/d$. In this case, we get a constant interaction strength, which  is in agreement with our phenomenological model.

One effect of electron-electron interaction is an enhancement of both the Land\'e $g$-factor and spin-orbit coupling, as discussed in Refs.~\cite{Janak1969, Zala2001, Chen1999}. By introducing the Fermi-liquid constants $F_0, F_1$ we can express the Land\'e $g$-factor enhancement as $\tilde{g}=g/(1-|F_0|)$. Spin-orbit coupling has an additional linear momentum dependence, which means that $|F_1|$ enters instead of $|F_0|$. Since $|F_0|>|F_1|$, this means that the $g$-factor will always be more strongly enhanced than the spin-orbit coupling. The enhancement is largest for large density of states, so that a strong confinement further enhances this effect.

\section{0.7 anomaly in BLG QPC}
\label{Sec:4}
As mentioned in the introduction, the 0.7 feature in the QPC conductance is seen as an additional kink below the lowest plateau sitting at around $0.7\times 2\, e^2/h$ \cite{Thomas1996} for systems without additional degeneracies of the spectrum. It has been observed and extensively studied in quantum point contacts in GaAs/AlGaAs heterostructures for both electrons \cite{Thomas1996, Thomas1998,Kristensen2000, Cronenwett2002, Reilly2002, Graham2003, Roche2004, DiCarlo2006, Sfigakis2008, Burke2012, Bauer2013, Iqbal2013, Brun2014, Iagallo2015, Smith2015, Smith2016}, and holes \cite{Rokhinson2006, Danneau2006b, Klochan2006, Danneau2008, Komijani2010, Klochan2011, Hudson2021}; signatures of the anomaly were also observed in Si/SiGe heterostructures \cite{Pock2016}.

This phenomenon cannot be explained within a single-particle picture \cite{Datta1995,Micolich2011}, and it is commonly admitted that it is directly linked to spin \cite{Thomas1996, Thomas1998}. In addition, this effect appears to be thermally activated and therefore not a ground-state property \cite{Kristensen2000}. Moreover, experiments show that the confinement potential seems to play a crucial in the strength and the position of this conductance feature \cite{Smith2015, Smith2016}. Recently, various explanations have been suggested  to capture the physical origin of the 0.7 anomaly, such as dynamical spin polarization or spin gap models due to electron-electron interaction \cite{Bruus2001, Kristensen2000, Pyshkin2000, Reilly2002, Reilly2005, Hudson2021}, Kondo effect \cite{Lindelof2001, Cronenwett2002, Meir2002, Hirose2003, Cornaglia2004, Rejec2006, Golub2006, Klochan2011}, Wigner crystallization \cite{Spivak2000, Matveev2004, Matveev2004a}, or charge density waves \cite{Sushkov2001} To our knowledge, no comprehensive study of the interaction-induced 0.7 anomaly in systems where both spin and valley degrees of freedom are degenerate has been reported so far.

In contrast to the conductance quantization, the shoulder-like feature appearing in the conductance cannot be explained by the non-interacting theory presented in Sec.~\ref{Sec:3}. In this section, we explore the possibility of explaining this special feature in the context of the interaction-induced $0.7$ conductance anomaly. As already discussed above, several microscopic theories were used to describe the $0.7$ anomaly.
Here, we do not specify any microscopic mechanism behind the anomaly, but, instead,  just assume that there is some that effectively leads to spin and/or valley polarization. 
Based on this assumption, we extend the phenomenological model of Ref.~\cite{Bruus2001} to four bands and the BLG band structure. The required polarization does not have to be static, it can fluctuate slowly compared to the typical traveling time through the constriction. For simplicity we nevertheless describe the model for a static situation. The ``classic'' $0.7$ effect is only seen in the lowest conductance step, Fig.~\ref{fig:exp}, so that below we restrict our consideration to the lowest size-quantized band shown in Fig.~\ref{fig:zoom}.

\subsection{Phenomenological model}
Following the general idea of \cite{Bruus2001}, we again use the Landauer-B{\"u}ttiker formula (\ref{eq:LandauerB})
for the conductance, here for the quantized band $N=0$ from Eq.~(\ref{eq:BLGQuantDisp}). The $0.7$ effect requires a finite temperature. Assuming that the energy scale for the variation of the transmission probability is smaller than that of the thermal distribution function, we approximate the former as a step function. A spin-valley subband contributes to the conductance as soon as the chemical potential reaches its lower band edge $\epsilon_{\sigma,\xi}^{0}$ within the temperature window.

In this section, we develop a phenomenological model to describe how interaction effects may influence these lower band edges beyond the self-consistent screening. There are two ways in which these can differ from the non-interacting single-particle ones. The first one is the spontaneous polarization mentioned above, which is assumed to be arbitrary in the space of spin and valleys. Already when the chemical potential is way below any of the relevant subbands, these subbands may be spontaneously split to different values of energy. The arrangement of these values, which are acquired for very low chemical potential and zero magnetic field, is referred to as the initial subband configuration. All subbands that are above the lowest subband are called minority bands; those that are characterized by the lowest band edge are majority bands. 

The second effect is the dependence of the subbands on the chemical potential when it is close to the band edge. A particular type of this dependence---pinning of the band edge to the chemical potential---gives rise to additional plateaus in the conductance. It is this interaction-induced dependence of the lower band edge of minority bands on the chemical potential that our phenomenological model describes for any assumed initial configuration. We then consider the corresponding evolution of the conductance with increasing in-plane magnetic field, and, by comparing the resulting behavior with the experimentally observed one (Sec.~\ref{Sec:2}), infer the initial splitting configuration.

\begin{figure*}
	\centering
	\includegraphics[width=1
	\textwidth]{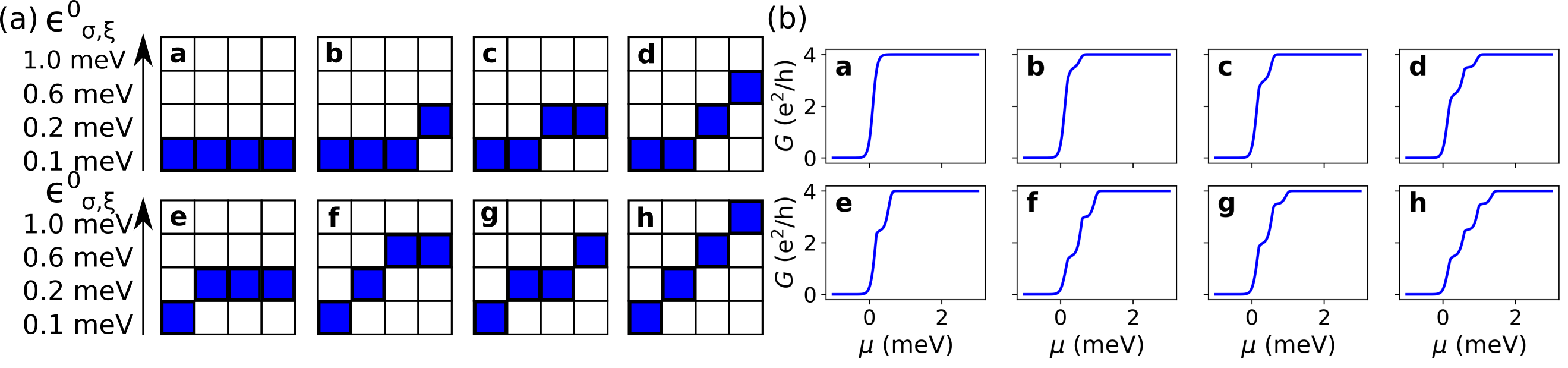}
	\caption{(a) Subfigures \textbf{a} to \textbf{h} show every possible initial subband structure. The $x$ axis labels spin and valley degree of freedom and the $y$ axis shows the position of the lower band edges in energy space with the values used for all plots. The assignment of spin $\sigma$ and valley $\xi$ to the $x$ axis is not unique, yet.  In \textbf{a} there is no initial splitting, all lower band edges are equivalent. In case \textbf{b}, three bands are fixed and there is one minority band. Whether the single minority band is spin up/down or valley $K$/$K^{\prime}$ can only be distinguished once we include a magnetic field. In \textbf{c} to \textbf{d} and \textbf{e} to \textbf{h} there are  two and three split subbands, respectively.  (b) Exemplary conductance for the initial configuration shown in (a) for zero magnetic field. The parameters are chosen to suppress the plateaus at multiples of $1\,e^2/h$, where the majority band is filled and minority bands are not yet populated. The majority bands are fixed at $0.1$ meV, the next levels correspond to  $\epsilon^0_1=0.2$\,meV, $\epsilon^0_2=0.6$\,meV, $\epsilon^0_3=1$\,meV. The temperature is $T=0.7$\,K and $C=6$\,meV$^{-3}$ for all minority bands [see Eq.~(\ref{eq:C})].}
	\label{fig:initial}
\end{figure*}

\subsubsection{General four-band model}

For a system with four degrees of freedom, like BLG, we label the subbands by their spin and valley index, i.e.,  $\epsilon_{\sigma,\xi}^{0}$. Moreover, we assume that the lower band edges of minority bands start to depend on the chemical potential once it reaches a certain value $\mu_{\sigma,\xi}$, i.e., $\epsilon_{\sigma,\xi}^{0}=\epsilon_{\sigma,\xi}^{0}(\mu)$. 

All possible initial spontaneously polarized configurations of the band edges are shown in Fig.~\ref{fig:initial}(a). 
For the analysis of the dependence of the band edges on $\mu$, we look at the local spin-valley energy-density functional in the form
\begin{align}
\mathcal{F}=\mathcal{E}[\{n_{\sigma,\xi}\}]-\mu\sum_{\sigma,\xi}n_{\sigma,\xi}.
\end{align}
Here, $\mathcal{F}$ is the free energy of the system and $\mathcal{E}$ is its internal energy. A diagrammatic approach to obtaining such a free energy and a corresponding analysis of possible instabilities in models with multiple species of quasiparticles is discussed, e.g., in Refs.~\cite{Miserev2019, Miserev2021}.

The lowest bands in Fig.~\ref{fig:initial} are majority bands with a fixed lower band edge and we decompose their density into $n=n_0+\delta n$. All changes with the chemical potential are included in $\delta n$. For minority bands, we do not make this decomposition, but assume that $n=0$ for $\mu<\mu_{\sigma,\xi}$. We approximate the free energy functional $\mathcal{F}$ as bilinear in all partial majority density contributions $\delta n$ and minority densities $n$, i.e.
\begin{align}
\mathcal{F}=
\sum_{i=\sigma,\xi}\left[(\alpha_{i}-\mu)n_{i}
+\frac{\beta_{i}}{2}n_{i}^{2}\ 
+\! \sum_{j=\sigma^{\prime},\xi^{\prime}} 
\gamma n_{i} n_{j}\right],
\label{eq:free-energy}
\end{align}
where  $\alpha_{\sigma,\xi}$, $\beta_{\sigma,\xi}$, and $\gamma$ are phenomenological constants to be determined experimentally and $n_{\sigma,\xi}$ is understood as $\delta n_{\sigma,\xi}$ for majority bands. The minimum of the energy functional is achieved when
\begin{align}
\frac{\partial \mathcal{F}}{\partial n_{\sigma,\xi}}=
\alpha_{\sigma,\xi}-\mu+\beta_{\sigma,\xi} n_{\sigma,\xi}+\gamma\sum_{\sigma^{\prime},\xi^{\prime}} n_{\sigma^{\prime},\xi^{\prime}}=0,
\label{eq:min-free}
\end{align}
which leads to solutions of the form \begin{equation}
    n_{\sigma,\xi}\propto (\mu-\mu_{\sigma,\xi}) \label{eq:nprop}
\end{equation}
for minority bands, where $\mu_{\sigma,\xi}$ is the critical chemical potential of the minority subband that depends on the parameters of the free energy Eq. (\ref{eq:free-energy}).

\subsubsection{Application to BLG}

At this point, we have to specify the band structure in order to get access to the 
single-particle densities entering the energy functional (\ref{eq:free-energy}).
For this purpose, we use the results of Sec.~\ref{Sec:IIIB} for the non-interacting 1D dispersion in the BLG QPC, and modify them to include the interaction effects at the phenomenological level.

Without a magnetic field, we consider a quartic dispersion relation of the form
\begin{align}
\epsilon_{\sigma,\xi}(k)=a k^4+\epsilon_{\sigma,\xi}^{0}(\mu),
\label{eq:DispLowBand}
\end{align}
where $a$ is a constant.
This form of the effective---modified by interactions---form of single-particle energies in BLG QPC is based on the fourth-order expansion of dispersion relation (\ref{eq:BLGQuantDisp}) for $N=0$, namely
\begin{align}
E_{N=0,\sigma}(k)\approx\left(\frac{\hbar^2}{2m}\right)^2\frac{k^4}{U} +\frac{U}{2}.
\end{align}

As shown above, the gap magnitude $U$ depends on the chemical potential through the self-consistent electrostatic screening. Specifically, $U=U(n)$, according to Eq. ~(\ref{eq:asymmetry}) with additional effects of the split gates, and $n=n(\mu)$ according to Eq.~(\ref{eq:densitypermu}). The chemical potential is set by the 2D density according to Eq.~(\ref{eq:2Ddensity}) and Eq.~(\ref{eq:electrodensity}). Within the lowest band, these dependencies are very smooth and do not lead to any additional features. The main role in our consideration is played by the interaction-induced bandgap that determines the band edge $\epsilon_{\sigma,\xi}^{0}(\mu)$. For this reason, we neglect all electrostatic contributions [effectively fixing $U=U(\mu_{\sigma,\xi})$] and introduce a new bandgap instead of $U/2$. One could just as well assume that this bandgap is applied on top of the fixed gap $U(\mu_{\sigma,\xi})/2$, since this would only lead to an overall shift and a redefinition of the origin. 

For the dispersion (\ref{eq:DispLowBand}), we get a one-dimensional density in the form
\begin{align}
n_{\sigma,\xi}(\mu)=\frac{2}{\pi} 
\left\{\frac{1}{a}\left[\epsilon_{\sigma,\xi}^{0}(\mu)-\mu\right]\right\}^{1/4}.
\end{align}
By combining this with Eq. (\ref{eq:nprop}), we thus get the dependence
\begin{align}
\epsilon_{\sigma,\xi}^{0}(\mu)=
\begin{cases}
\mu-C_{\sigma,\xi}(\mu-\mu_{\sigma,\xi})^{4},&\mu>\mu_{\sigma,\xi}\\
\mu_{\sigma,\xi}, &\mu<\mu_{\sigma,\xi}
\end{cases}.
\label{eq:C}
\end{align}
where $C_{\sigma,\xi}$ is a phenomenological constant depending on the parameter $a$ as well as the parameters of the free energy functional $\mathcal{F}$. This means that once the chemical potential reaches the lower band edge of a minority band they become pinned together over a certain energy range. For continuity reasons we require $\epsilon_{\sigma,\xi}^{0}= \mu_{\sigma,\xi}$, i.e., the initial configuration determines the critical chemical potentials.
It is worth emphasizing that the enhanced density of states at the bottom of the almost flat (quartic in momentum) band in BLG QPC (\ref{eq:DispLowBand}) is expected to enhance the role of interactions compared to the case of conventional parabolic bands. 

\subsubsection{Resulting conductance}

With the step-function transmission probabilities, the conductance reads:
\begin{align}
G(T)=\frac{e^2}{h}\sum_{\sigma,\xi}f(\epsilon_{\sigma,\xi}^{0}(\mu)-\mu).
\end{align}
At this point, one should note that the Fermi function $f(x)$ is close to $1/2$ for $|x|\ll k_B T$. For fixed lower band edges, this corresponds to a very small region and does not lead to conductance anomalies, but for the pinned band edges and finite temperature considered here it does. If we tune the chemical potential through all band edges, the crossing of a fixed majority band corresponds to a plateau of $1\,e^2/h$, while for every minority band we get an additional less sharp one at $0.5\,e^2/h$. Any additional plateau from a majority band at $1\,e^2/h$ can be smeared by temperature. If we have several minority band edges at different initial energies,
the distance between the bands compared to the temperature determines whether lower minority bands already contribute fully or not, cf. Ref.~\cite{Bruus2001}.

The conductance corresponding to the initial splitting configurations from Fig.~\ref{fig:initial}(a) is shown in Fig.~\ref{fig:initial}(b).
The values of the additional shoulders are summarized in  Table~\ref{tab:cond}.
In experimental conductance curves in Fig.~\ref{fig:exp}(d), there is one additional shoulder at around $2.5\,e^2/h$ and another one around  $3.5\,e^2/h$ for zero magnetic field. Thus we can rule out case \textbf{a}, because it does not have any additional shoulder and cases \textbf{f}, \textbf{g}, and \textbf{h}, which have a too low first shoulder from the very beginning. Only cases \textbf{b}, \textbf{c},\textbf{d} and \textbf{e} from Fig.~\ref{fig:initial} are relevant here. 
\begin{table}
\addtolength{\tabcolsep}{-0pt}
    \begin{ruledtabular}
    \begin{tabular}{m{0.2\columnwidth}m{0.8\columnwidth}}
    
    case &  conductance shoulders in units of $e^2/h$  \\ 
 \hline 
 \textbf{a} & none \\ 
 \hline 
 \textbf{b} & 3.5 \\ 
 \hline 
 \textbf{c} & 3 \\ 
 \hline 
  \textbf{d} & 2.5 and 3.5\\ 
 \hline 
 \textbf{e} & 2.5 \\ 
 \hline 
 \textbf{f} & 1.5 and 3 \\ 
 \hline 
 \textbf{g} &2 and 3.5 \\ 
 \hline 
 \vspace{0.05 cm}\textbf{h} & 1.5 and 2.5 and 3.5 
    \end{tabular}
    \caption{Values of conductance shoulders for initial configurations shown in Fig.\,\ref{fig:initial}\label{tab:cond}}
    \vspace{-0.05 cm}
    \end{ruledtabular}
\end{table}

\begin{table}
    \begin{ruledtabular}
    \begin{tabular}{m{0.2\columnwidth}m{0.8\columnwidth}}
    
 case & spin configurations \\ 
 \hline 
 \textbf{a} & $(\uparrow \uparrow \downarrow \downarrow)$ \\ 
 \hline 
 \textbf{b} & $(\uparrow \uparrow \downarrow) \downarrow,\quad(\uparrow \downarrow \downarrow)\uparrow$ \\ 
 \hline 
 \textbf{c} & $\uparrow \uparrow \downarrow \downarrow,\quad(\uparrow \downarrow) (\downarrow\uparrow),\quad  \downarrow \downarrow\uparrow \uparrow$ \\ 
 \hline 
  \textbf{d} & $\uparrow \uparrow \downarrow \downarrow,\quad(\uparrow \downarrow) \downarrow \uparrow,\quad(\uparrow \downarrow) \uparrow \downarrow,\quad  \downarrow \downarrow\uparrow \uparrow$\\ 
 \hline 
 e & $\downarrow(\downarrow\uparrow\uparrow),\quad \uparrow(\uparrow\downarrow\downarrow)$ \\ 
 \hline 
 \textbf{f} & $\uparrow \uparrow\downarrow \downarrow,\quad \downarrow \uparrow(\uparrow \downarrow), \quad \uparrow \downarrow(\uparrow \downarrow), \quad  \downarrow \downarrow\uparrow \uparrow$ \\ 
 \hline 
 \textbf{g }& $\uparrow\downarrow\downarrow\uparrow,\quad \downarrow\uparrow\uparrow\downarrow,\quad \downarrow(\uparrow\downarrow)\uparrow,\quad \uparrow(\uparrow\downarrow)\downarrow$ \\ 
 \hline 
 \vspace{0.05 cm}\textbf{h} & $\uparrow\uparrow\downarrow\downarrow,\quad \uparrow\downarrow\downarrow\uparrow,\quad \uparrow\downarrow\uparrow\downarrow,\quad \downarrow\downarrow\uparrow\uparrow,\quad \downarrow\uparrow\uparrow\downarrow,\quad \downarrow\uparrow\downarrow\uparrow$ \\
 \end{tabular}  
 \caption{Different spin configurations for the arrangements shown in Fig.~\ref{fig:initial}; brackets ``$(\ldots)$'' denote all possible permutations. The order corresponds to the subband ordering shown in Fig.~\ref{fig:initial}. For each permutation we can assign the valley indices in four different ways, which does not lead to different behavior in this model and is thus not distinguishable. For example, the last arrangement in case g corresponds to the eight valley-resolved cases: ${\uparrow(\uparrow\downarrow)\downarrow}=\lbrace {\uparrow_1\uparrow_2\downarrow_1\downarrow_2,}
  \ {\uparrow_1\uparrow_2\downarrow_2\downarrow_1,}\ {\uparrow_2\uparrow_1\downarrow_1\downarrow_2,}\ {\uparrow_2\uparrow_1\downarrow_2\downarrow_1,}\ {\uparrow_1\downarrow_1\uparrow_2\downarrow_2,}\ {\uparrow_1\downarrow_2\uparrow_2\downarrow_1,}
  $ 
  ${\uparrow_2\downarrow_1\uparrow_1\downarrow_2,}
  \ {\uparrow_2\downarrow_2\uparrow_1\downarrow_1}
  \rbrace$, where the index $1,2$ corresponds to the $K,K^{\prime}$ valley, respectively.
  \label{tab:assignments}}
    \end{ruledtabular}
\end{table}

\subsection{Behavior of conductance in magnetic field}

In order to distinguish between the cases of spin, valley or spin-valley splitting, we consider the behavior of the conductance in parallel magnetic field $B$.
This is incorporated by the following replacement
\begin{align}
\epsilon_{\uparrow,\xi}^{0}&\rightarrow \epsilon_{\uparrow,\xi}^{0}- \frac{1}{2}g\mu_B |B|, \label{symsplit1}\\
\epsilon_{\downarrow,\xi}^{0}&\rightarrow
\epsilon_{\downarrow,\xi}^{0}+ \frac{1}{2}g\mu_B |B|.\label{symsplit2}
\end{align}

There are in total 6 possibilities of assigning $2\times 2$ spins to four subbands. Since one cannot distinguish different valley indices this way, after this assigning, the spins are still four-fold degenerate in their valley index. We only know that each valley has two opposite spins. The permutation of spins within subbands with the same lower band edge does not change the outcome. From this we get in total 26 different arrangements with distinct development with magnetic field for the eight initial cases shown in Fig.~\ref{fig:initial}. These are shown in Table~\ref{tab:assignments}.

\begin{figure*}
	\centering
	\includegraphics[width=1
	\textwidth]{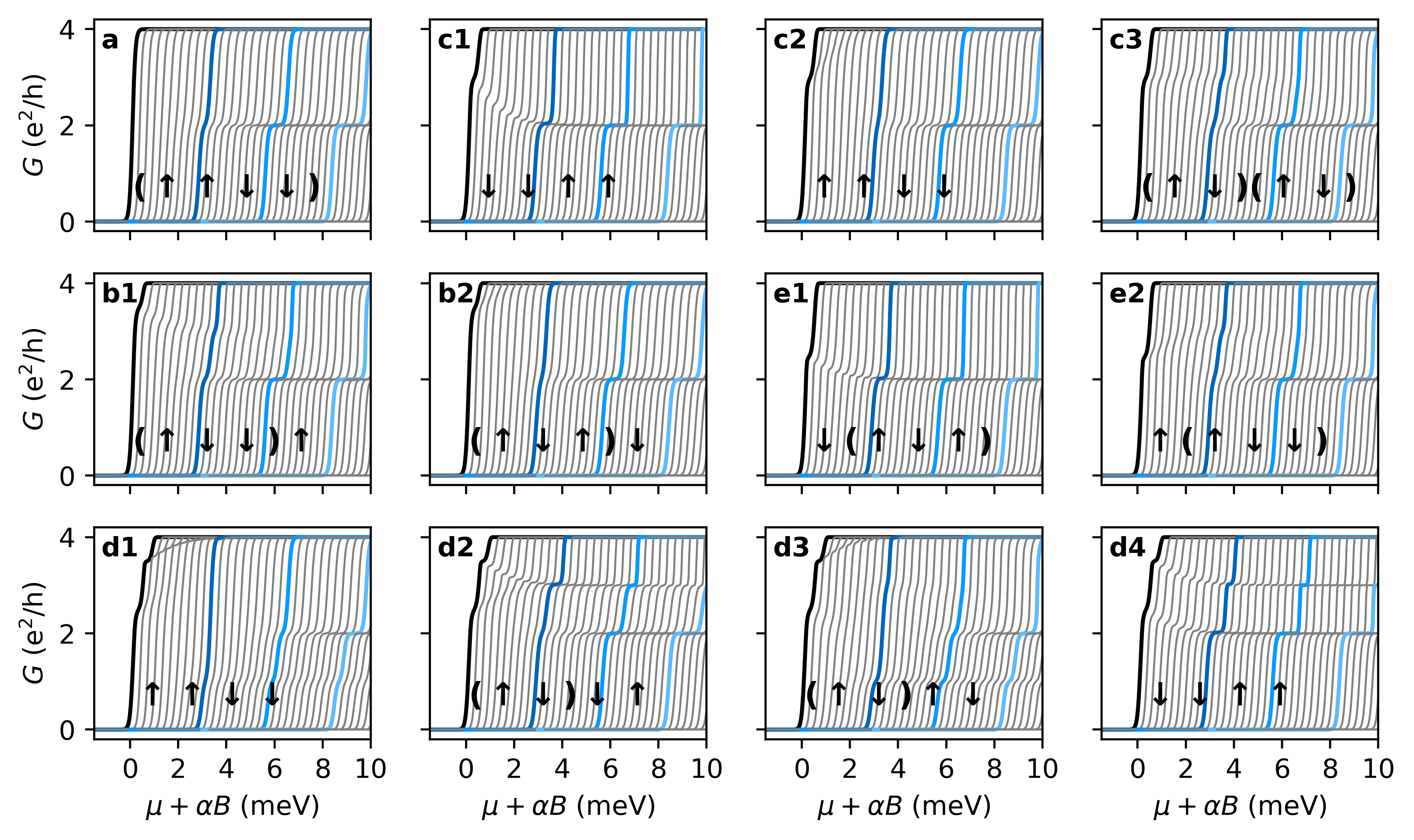}
	\caption{Behavior of the four different initial splitting cases \textbf{a}, \textbf{b}, \textbf{c} and \textbf{e} from Fig.~\ref{fig:initial} in an in-plane magnetic field. Conductance curves for magnetic fields between $0$T and $8$T are shown with a horizontal shift parametrized by $\alpha=1.5\mathrm{meV}/\mathrm{T}$. The thick black curve corresponding to $B=0$\,T is non-shifted. The blue lines correspond to $2$\,T ,  $4$\,T and and  $6$\,T, as in Fig.~\ref{fig:zoom}. Without any initial splitting, there is no continuous development of a shoulder in \textbf{a}, the additional plateau appears, as soon as it can be resolved. The particular assignment of spin to the subbands is irrelevant: all six possibilities are indistinguishable. In case \textbf{c}, three different scenarios are possible. Each case happens for all four possible valley assignments. For cases \textbf{b} and \textbf{e}, there are two distinguishable spin configurations; for case \textbf{d} four.  Same parameters as in Fig.~\ref{fig:initial}; according to the measured  Land\'e $g$-factor, $g$=4 was chosen. }
	\label{fig:magnetic}
\end{figure*}

The magnetic field behavior of the relevant cases \textbf{b}, \textbf{c}, \textbf{d}, \textbf{e}  is shown in Fig.~\ref{fig:magnetic} in analogy to Fig.~\ref{fig:zoom}(a). Here one should note, that the initial spontaneous splitting is a spontaneous symmetry breaking and if the magnetic field is tuned adiabatically, it will always favor the initial splitting in the direction of the magnetic field.  Behavior like case \textbf{c2}, where the initial spontaneous splitting is opposite to the Zeeman splitting will only be observed if the magnetic field is switched on very fast.

\begin{figure*}
	\centering
	\includegraphics[width=1
	\textwidth]{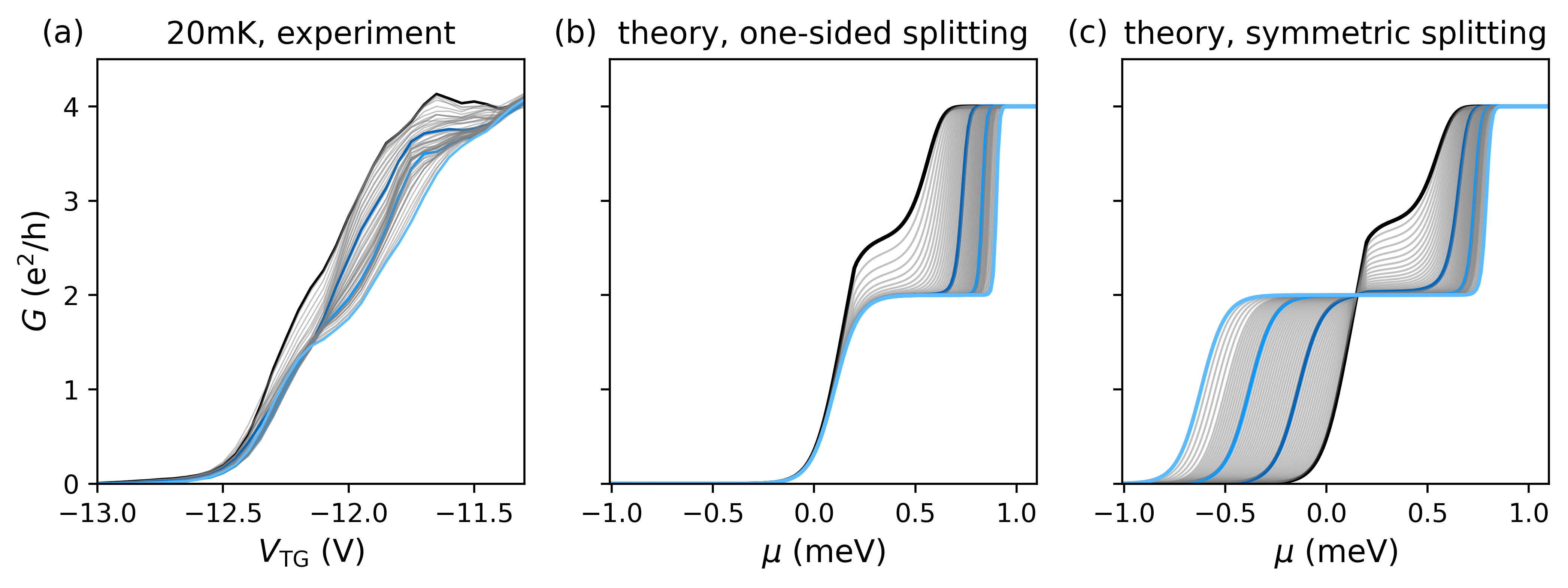}
	\caption{Zoom into the evolution of the lowest conductance step with increasing magnetic fields (without shifts). The colored curves correspond to those marked in Fig.~\ref{fig:exp}a. (a): Experimental data for $20$\,mK. Curves for higher magnetic fields show lower conductance at a fixed value of the top-gate voltage. (b): Theoretical curve for case \textbf{c1} in Fig.~\ref{fig:magnetic} with the non-symmetric splitting (sp.) introduced in Eqs.~(\ref{asymsplit1}) and (\ref{asymsplit2}). Same parameters as in Fig.~\ref{fig:initial}; according to the measured  Land\'e $g$-factor, $g$=4 was chosen. The behavior of (a) is qualitatively replicated. (c): Same as in (b), but with a symmetric Zeeman splitting introduced in Eqs.~(\ref{symsplit1}) and (\ref{symsplit2}). There is a fixed(crossing) point clearly absent in the experimental data. As is apparent by comparing the distance of the blue plateaus in chemical-potential space in (b) and (c) a symmetric splitting enhances the $g$-factor even more.}
	\label{fig:comp}
\end{figure*}

A comparison of the experimental data and theoretical ones for symmetric splitting Eqs.~(\ref{symsplit1}) and (\ref{symsplit2}) and a phenomenological asymmetric one, where we use
\begin{align}
\epsilon_{\uparrow,\xi}^{0}&\rightarrow \epsilon_{\uparrow,\xi}^{0}- g\mu_B |B|, \label{asymsplit1}\\
\epsilon_{\downarrow,\xi}^{0}&\rightarrow
\epsilon_{\downarrow,\xi}^{0},\label{asymsplit2}
\end{align}
assuming that the spin up band is energetically higher
  is shown in Fig.~\ref{fig:comp}. From this we see that an asymmetric splitting, Eqs.~(\ref{asymsplit1}) and (\ref{asymsplit2}),
yields a better agreement with the experimental observations in this particular case, that will turn out to be the most relevant one. However, owing to the special dependence of minority bands on the chemical potential, this asymmetric replacement rule may lead to unphysical half-integer plateaus in high magnetic fields for some initial configurations. Therefore, we have used the symmetrical splitting introduced in Eqs.~(\ref{symsplit1}) and (\ref{symsplit2}) to produce Fig.~\ref{fig:magnetic}. 

\section{Discsussion}
\label{Sec:5}
Let us now compare the results of our phenomenological model with the experimental results reported in Sec.~\ref{Sec:2}.
Many, but not all, features in Fig.~\ref{fig:exp}, e.g., conventional conductance quantization, can be explained without considering interaction effects. Other features, e.g., additional shoulders in the conductance curves and behavior of the $g$-factor, are compatible with the phenomenological model presented in Sec.~\ref{Sec:4}.

\subsection{Conductance plateaus}

Every time the chemical potential, tuned by the top gate voltage, reaches a new lower band edge, the conductance makes a step of $1\,e^2/h$ per spin and valley. For zero magnetic field, the plateaus are at multiples of $4\,e^2/h$, which can be clearly seen in the cadence plots in Fig.~\ref{fig:zoom}. This is in contrast to Ref.~\cite{Sakanashi2021}, where the valley splitting was observed in a similar setup with changing the split-gate voltage, but at much higher back-gate voltage.
The Zeeman coupling of the spin to the in-plane magnetic field leads to appearance of steps at multiples of $2\,e^2/h$ for higher magnetic field. The additional plateaus become visible when the Zeeman-split bands have a spacing that can be experimentally resolved, which occurs in our case above 2\,T. 

\subsection{Effective \textit{g}-factor}

The Zeeman splitting in the first three subbands shown in Fig.~\ref{fig:3}(a) for $20$mK is very close to linear at sufficiently high magnetic fields. For the lowest subband, a nearly constant splitting is observed up to nearly $4$T. The extracted $g$-factors show a strong enhancement compared to the bare value of $g=2$ for BLG. We attribute this enhancement to the strong confinement and interaction effects, similar to those discussed in Ref.~\cite{Menezes2017}.

These effects are strongest for the lowest subbands because of lower densities in the almost flat (quartic) band, which is consistent with the increase of the enhancement with lowering the band index. This effect should only be present for electrons going through the constriction. Electrons that bounce back and stay in the 2D region are at too high densities for the interaction-induced enhancement to be visible. This effect combined with the peculiar low-field behavior of the lowest subband strongly hints at the importance of interaction effects in this experiment.

\subsection{Additional features}

There are two additional features below the conventional conductance quantization step at $G=4\,e^2/h$, which thus only involve the lowest quantized subband. The first one starts at around $3\,e^2/h$ at zero magnetic field, and the second one at around $2.5\,e^2/h$.
\subsubsection{\textit{0.7} anomaly}
One might try to identify both these features with case \textbf{d} in Fig.~\ref{fig:initial}, where there are two additional shoulders at zero magnetic field. However, considering the conductance behavior in magnetic field shown in Fig.~\ref{fig:magnetic}, one sees that when these two additional shoulder move down with magnetic field, as in case \textbf{d4}, the additional plateau at $3\,e^2/h$ persists for higher magnetic fields. We do not see such a plateau in the experimental data. The feature starting around $2.5\,e^2/h$ in Fig.~\ref{fig:zoom}(a) moving down into the $2\,e^2/h$ plateau is visible in Fig.~\ref{fig:exp}(a) as a splitting of the spin-valley subbands at vanishing magnetic field, which makes it a strong contender for the $0.7$ effect. It also leads to the non-linear behaviour of the Zeeman splitting in the first subband in Fig.~\ref{fig:3}. 

Since we have ruled out case \textbf{d}, only the cases \textbf{b}, \textbf{c} and \textbf{e} in Table~\ref{tab:cond} are still possible. Upon comparing the magnetic field behavior with 
Fig.~\ref{fig:magnetic}, we conclude that we are in either case \textbf{c1}  or \textbf{e1}. While one might not be convinced by the value of the theoretical shoulder of case \textbf{c}, that is $3\,e^2/h$ compared to the experimental one at $2.5\,e^2/h$, which is exactly the value in case \textbf{e}, as shown in Fig.~\ref{fig:initial}, this would require the spin-up state of one valley split to the same energy as the non-spin split bands of the other valley, which implies an accidental fine-tuning. Instead, if only one valley was spin split, cases \textbf{d}, \textbf{f} and \textbf{g} would be way more probable, but these were already ruled out. Thus, we identify the experimentally observed behavior as case \textbf{c1}, which assumes an initial spin splitting, but no valley splitting.

It is also clear that, in contrast to \cite{Banszerus2020}, we do not see any crossing of Zeeman-split bands. A shoulder similar to ours but at $2\,e^2/h$ was attributed to the substrate-enhanced Kane-Mele spin-orbit coupling in Ref.~\cite{Banszerus2020}.
We note that such effects of the weak spin-orbit coupling can be observed only at very low temperatures, but we still see a similar effect at $4$K. Finally, the proposed Kane-Mele spin-orbit splitting would lead to opposite spin splitting in the two valleys, so that there is no net spin splitting, as detailed in Appendix \ref{sec:eff2DModel}. However, the observed Zeeman splitting at low magnetic fields suggests the presence of spontaneous net spin splitting in our case, while the enhancement of the effective $g$-factor points towards rather strong interaction effects. We thus identify this feature in the conductance as an interaction-induced $0.7$ anomaly. As mentioned in Ref. \cite{Heyder2015}, the exact value of the shoulder may depend on the exact QPC geometry, so that it may also appear very close to the value of $0.5\times 4\,e^2/h$.
 
\subsubsection{Fabry-P\'{e}rot resonances}

We identify the upper feature in the lowest subband conductance at low magnetic field, corresponding to an additional peak in the low-temperature plot in Fig.~\ref{fig:zoom}(c) at around $-11.8$\,V that goes vertically through the right spin-split band, as a Fabry-P\'{e}rot resonance \cite{Young2009, Shytov2008, Rickhaus2013, Varlet2014, Varlet2016, Du2018, Kraft2020}. 
In Fig.~\ref{fig:exp}(e), this additional feature is seen as a faint bright curve moving down from the $4\,e^2/h$ plateau (at weak fields) and merging with the 0.7-feature to from the spin-split $2\,e^2/h$ plateau at magnetic field around $4$\,T. Note that at this same value of magnetic field, the Zeeman splitting of the lowest subband starts growing linearly with magnetic field, see Fig.~\ref{fig:3}(a). 
With increasing temperature, this feature disappears, in contrast to the 0.7 anomaly, see Fig.~\ref{fig:zoom}(d).

The Fabry-P\'{e}rot resonances in our geometry emerge from interferences of electronic waves in the 2D region, which are back-scattered from the interface with the contact on one hand, and the barrier created by the split gates, see Ref.~\cite{Kraft2018b}.  
In a parallel magnetic field, there are two Fermi-wavevectors, one for each spin, so that the minima and maxima of these oscillations disperse with the magnetic field. Since Fabry-P\'{e}rot resonances correspond to electrons bouncing back and forth between the contacts and the split gate, these electrons are inherently two-dimensional and are not affected by the enhancement of the $g$-factor in the QPC region.  
A closer look reveals additional Fabry-P\'{e}rot peaks at other values of the top-gate voltage, which do not move to different plateaus in the considered range of magnetic fields. For a discussion of this effect, see App.~\ref{app:FP}.

\section{Conclusions}
\label{Sec:6}
In conclusion, we have studied an electrostatically defined QPC in BLG which shows a zero field quantized conductance in steps of $4\,e^2/h$ owing to the spin and valley degeneracy. In an in-plane magnetic field, a splitting of the first three subbands at $20$\,mK is observed that results from the Zeeman spin splitting, while the valley degeneracy is not affected. Additionally, a $0.7$-like structure is located below the lowest size-quantized energy level which develops into the lowest spin split subband at $2\,e^2/h$. This additional feature is also observed in the $4$K data, where only the splitting of the lowest band is clearly resolved. On top of the quantized conductance we observe Fabry-P\'{e}rot resonances. Because of the higher densities in the 2D region and the relatively small bare Zeeman splitting in BLG, these stay at fixed top gate values with increasing magnetic field. 

From the Zeeman energy splitting, the effective 1D $g$-factors in an in-plane magnetic field are found to be increasingly enhanced for lower subbands compared to the bare 2D Land\'e $g$-factor $g=2$ in BLG. Moreover, the fact that the linear fitting of the Zeeman energy splitting for $N=0$ does not extrapolate to zero at $B=0$ further indicates the spontaneous spin polarization of the lowest subband. The behavior of the Zeeman spitting is a clear sign of the importance of interaction effects and confinement in this experiment. Based on this, we also attribute the observed shoulder below the lowest subband to the $0.7$ anomaly stemming from the interaction-induced lifting of the band degeneracy. 

We employ a phenomenological model to qualitatively describe the behavior of this feature in the applied in-plane magnetic field. In this model we assume, that each spin-valley subband can be spontaneously split by the electron-electron correlations. By comparing the development of resulting features in a magnetic field, Fig.~\ref{fig:magnetic}, with the experimental conductance curves in Fig.~\ref{fig:zoom}, we conclude, that the observed behavior can be explained by the assumption of an effective spontaneous spin splitting, while the valley degree of freedom is not affected. This is in a full agreement with the picture of spontaneous spin polarization inferred from the measured Zeeman splitting. 

Our experimental findings, supported by phenomenological calculations and combined with those of Ref.~\cite{Kraft2018b} for out-of-plane magnetic field, establish the exquisite tunability of spin and valley degree of freedom by the application of gates or external magnetic fields. Furthermore, our results also demonstrate relevance of electron-electron correlations in BLG QPC geometry, as well as a possibility to control the effective strength of interactions by means of electrostatic spatial confinement by a combination of external gates. 

Apart from developing the microscopic theory of 0.7 anomaly in BLG QPC, several questions regarding the phenomenological model still remain. While it is straightforward to obtain a free energy of the form of Eq.~(\ref{eq:free-energy}) for a 2D model with quadratic band structure (cf. Ref.~\cite{Miserev2019}), the corresponding calculation for a hybrid 2D-1D geometry is much more involved. Within such a derivation, it would be especially interesting to show the microscopic origin of the parameters of the phenomenological description. In particular, such a calculation would yield their explicit dependence on the applied magnetic field.

There are several additional ingredients that could be combined with this sort of setup.
In particular, in Ref.~\cite{Gold2021} it was observed that, at least if there is a substantial gap and the trigonal warping is relevant, electrons might predominately orient along the lattice directions and not take the shortest path, which is expected to affect the conductance of the QPC in the corresponding regime of gate voltages.
Further, while the intrinsic spin-orbit coupling in BLG is very weak, using an additional layer with strong spin-orbit coupling, e.g., a layer of a transition metal dichalcogenide, should induce noticeable proximity spin-orbit related effects \cite{Wang2015,Thomson2021} and may lead to topologically nontrivial states. 
In addition, the introduction of a finite twist between the layers may also lead, at certain fillings and twist angles, to topological states \cite{Cao2018,Thomson2021}. In order to open a gap in such a system, spin-orbit coupling has to be added as well. To what extent these states can be manipulated with gates and external magnetic fields and what role interaction effects play in such engineered sample are questions worth  exploring. The analysis of the present paper serves as the starting point for further studies in this direction.

\section*{Acknowledgments}
We thank Alexander Dmitriev, Angelika Knothe, Ralph Krupke, Alex Levchenko, Christoph Stampfer, and Xavier Waintal for discussions and comments. This work was partly supported by Helmholtz
society through program STN, by the DFG via the projects
DA 1280/3-1 and GO 1405/5 within FLAGERA Joint Transnational Call 2017 (Project GRANSPORT), and by the RFBR  (grant No. 20-02-00490).

\appendix

\begin{figure}
\centering
	\includegraphics[width=\columnwidth]{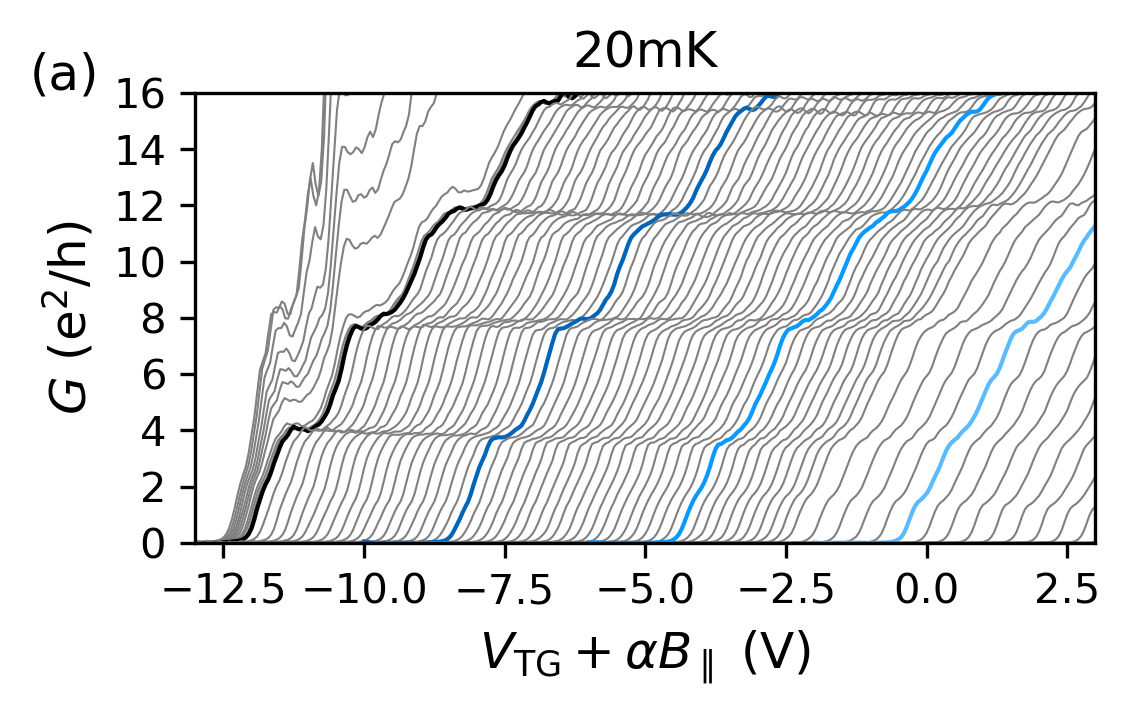}
		\includegraphics[width=
	\columnwidth]{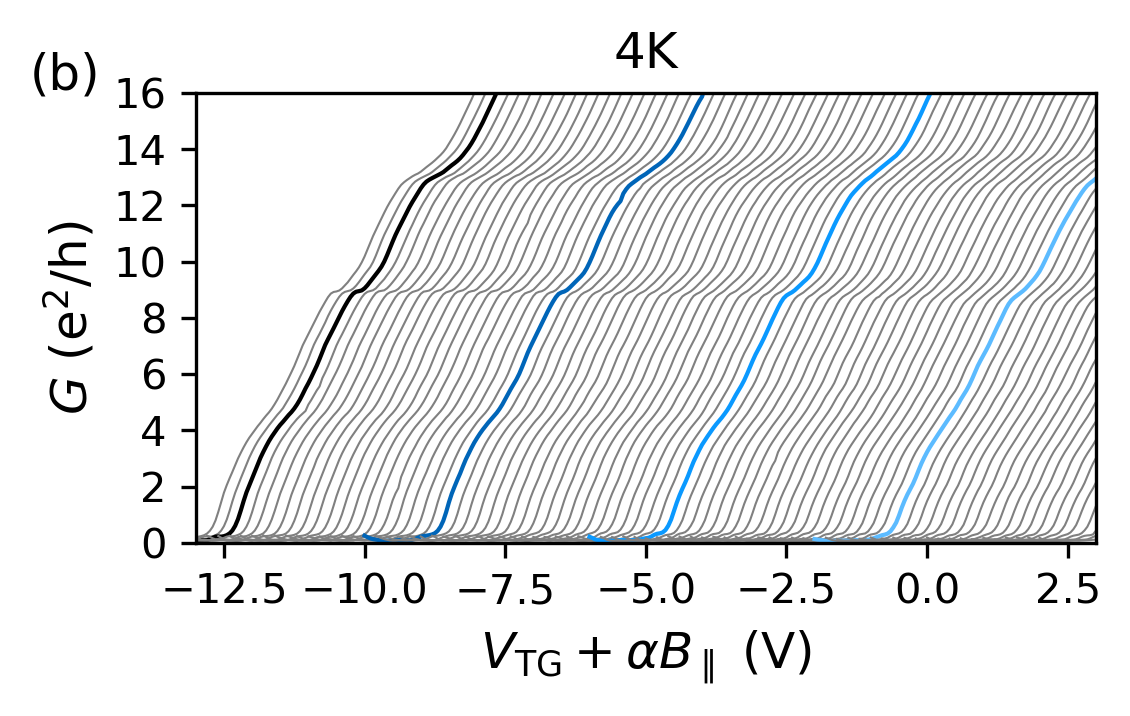}
	\caption{(a) and (b): Differential conductance $G$ as a function of the top gate voltage $V_\mathrm{TG}$ for different values of magnetic field $B$ and for $20$\, mK and $4$\, K, respectively. The curves are shifted with $\alpha = 2{\mathrm{V}}/{\mathrm{T}}$. Figures \ref{fig:zoom}(a) and \ref{fig:zoom}(d) are a zoom into the lowest plateau shown here. }
	\label{fig:expadd}
\end{figure}
\section{Additional plots \label{sec:addPlots}}

The $0.7$ anomaly discussed in the main text is only visible in the lowest conductance plateau. The other discussed features, including the Fabry-P\'{e}rot resonances, are visible in other plateaus as well. In this Appendix, we show cadence plots of the measured conductance for larger ranges of conductances and top-gate voltages, than in  Figs.~\ref{fig:zoom}(a) and \ref{fig:zoom}(b). Figures \ref{fig:expadd}(a) and \ref{fig:expadd}(b) show the differential conductance as a function of the applied top-gate voltage with horizontal shift linear in the applied in-plane magnetic field. The additional $0.7$-shoulder is seen in the lowest step for both temperatures. The main difference between the two temperatures is the smoother and flatter behavior for higher temperatures. Moreover, there are two additional features that are only visible in the $20$\,mK case, Fig.~\ref{fig:expadd}(a). For magnetic fields below $0.2$T, the aluminum leads are still superconducting, so that the conductance is affected by superconducting fluctuations. Additionally, one sees Fabry-P\'{e}rot resonances, which are most clear on top of plateaus.

Since the aluminum leads are superconducting at $20$ mK, a finite magnetic field is needed to kill this effect and curves below $0.2$T show a higher conductance than the quantized values. This should be contrasted with the data shown in Fig.~\ref{fig:zoom}(b) for $4$K, where there are no superconducting effects even at vanishing magnetic field.
The superconducting proximity effect for the QPC in BLG is out of scope of the present paper; the analysis of conductance curves affected by superconducting fluctuations is an interesting task from both the experimental and theoretical points of view (for a related analysis of the supercurrent in this geometry, see Ref.~\cite{Kraft2018a}). 

\begin{figure}[htp]
	\centering
\includegraphics[width=0.95
\linewidth]{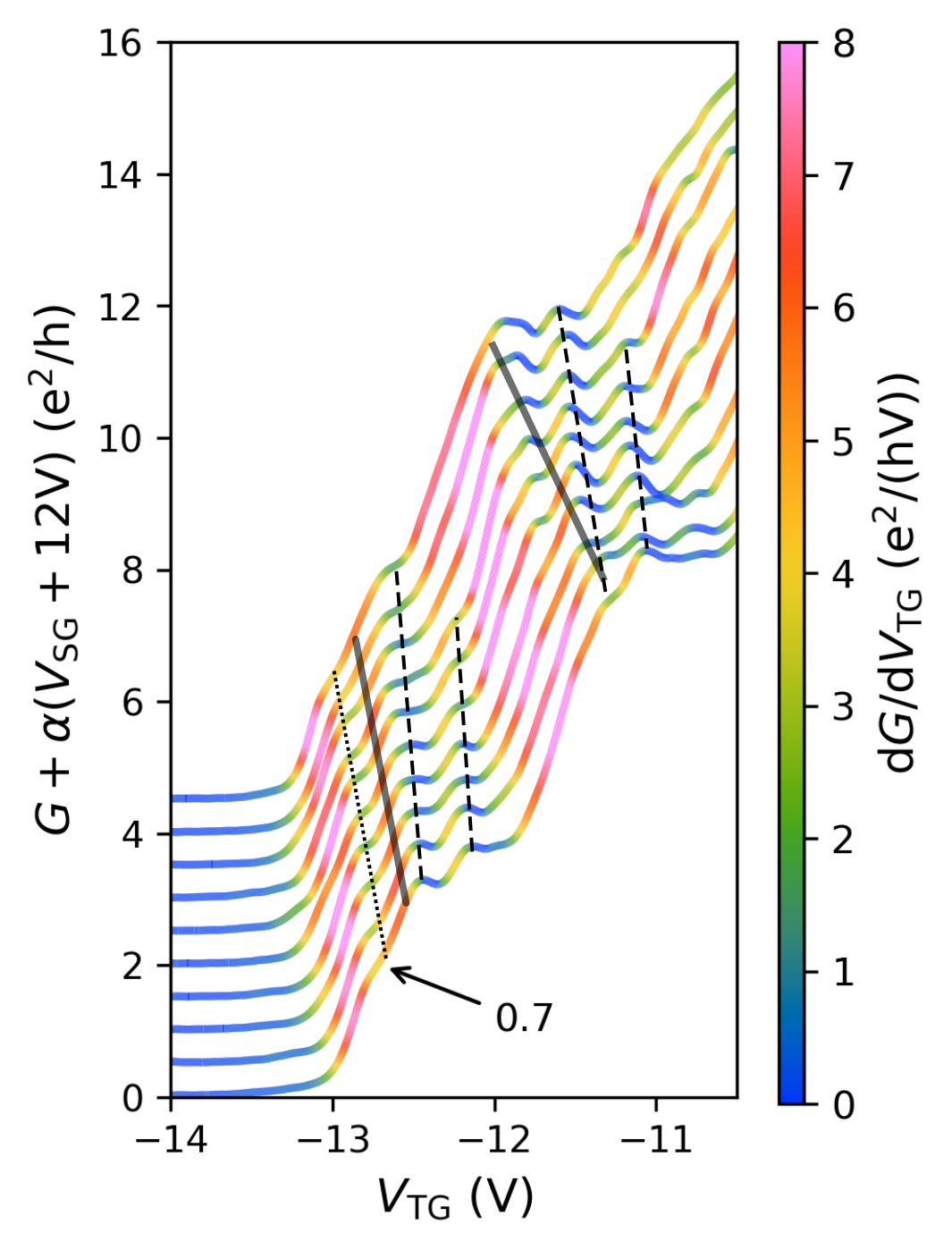}
	\caption{ Cubic spline fit of the differential conductance $G$ of the same sample in a different cooldown for $V_\text{BG}=10$~V and a perpendicular magnetic field of $20$~mT as a function of the $V_\mathrm{TG}$ in elevating $V_\text{SG}$ for $20$\,mK. The curves are shifted vertically with $\alpha = 10{e^2}/{h\mathrm{V}}$ and colored according to their first derivative.  The onset of the conductance plateau (black solid lines serving as guides for the eye) shows a clear dependence on the exact gate configuration. The black dotted line shows the dispersion of a feature, that we identify as a 0.7 shoulder, with a dispersion parallel to the onset of the plateau, in agreement with its quasi-1D nature. Fabry-P\'{e}rot oscillations are marked by black dashed lines and show a different (weaker) split-gate voltage dependence, since they are generated by the 2D lead modes.  }
	\label{fig:Variation}
\end{figure}

Since it is well known that the 0.7 anomaly is very sensitive to the exact shape of the constriction, we include data of the same sample in a different cool down at $T=20$ mK and with a perpendicular magnetic field of $20$ mT. The back-gate voltage is again $V_\text{BG}=10$~V and the split-gate voltage ranges between $-12$~V and $-11.5$~V.  Figure~\ref{fig:Variation}  shows a cubic spline fit of the obtained conductance data in a form similar to Fig.~\ref{fig:exp} (b) and (c) but the vertical shifts correspond to different split-gate voltages, starting at $V_\text{SG}=-12$~V for the lowest curve and ending with $V_\text{SG}=-11.5$~V as a function of $V_\text{TG}$. The curves are colored according to their derivative. The thick solid lines mark the onset of the conductance plateaus, showing their dependence on the exact confinement condition, i.e., the split-gate voltage. The lowest curve corresponds to the same split-gate and back-gate configuration as the data in the main text, where we have identified the 0.7 anomaly by its magnetic field dependence, see Fig.~\ref{fig:exp} (b) and (e). 

In this cooldown, we see a similar feature, marked by the arrow. When following the split-gate dependence of this feature (black dotted line), one observes that it stays parallel to the onset of the lowest plateau, which verifies that it is a feature of the QPC modes. Additionally, we again see Fabry-P\'{e}rot oscillation on top of the $4\,e^2/h$ and $8\,e^2/h$ plateaus (black dashed lines). Since they are generated by the lead modes, they show a different dispersion with the split-gate voltage. They always appear at the same electronic 2D density, which is only slightly tuned by the split-gate voltage. The onset of the conductance steps (and the 0.7 anomaly) is much more strongly dependent on the exact gate configuration, which makes the two effects clearly distinct.


\section{Fabry-P\'{e}rot resonances \label{app:FP}}

In our experiment (Fig.~\ref{fig:exp}), the 0.7 shoulder in the conductance of the lowest subband merged at magnetic field of about $4$\,T merges with an  additional conductance feature, which was identified with the Fabry-P\'{e}rot resonances in the main text. Here, we present additional details supporting this identification.
The transmission coefficient accounting for the Fabry-P{\'e}rot resonance can be described by
\begin{align}
\mathcal{T}=\frac{1}{1+F\sin^2\left[L k \cos(\theta) \right]}, 
\label{eq:FabryTrans}
\end{align}
where $F$ is the finesse and $\theta$ is the angle of incidence of the electron wave. 
At very low temperatures, the contribution of the resonance to the conductance is given by
\begin{align}
G = \frac{e^2}{h}\frac{1}{1+F \sin^2\left(L k_F  \right)}. \label{eq:FP}
\end{align}

\begin{figure}[t!]
	\centering
	\includegraphics[width=.95\linewidth]{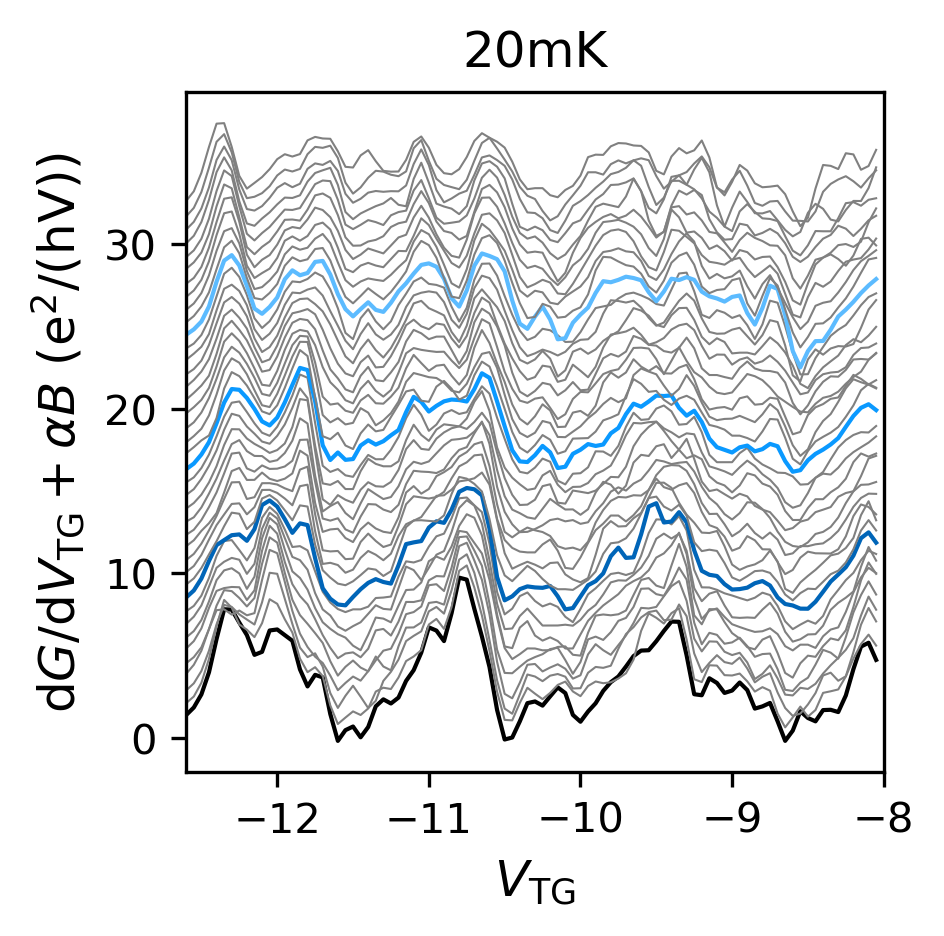}
\caption{Differentiated differential conductance at $20$\,mK for different magnetic fields with a vertical shift with $\alpha=4\mathrm{e^2/(hVT)}$. Only magnetic fields with magnitudes of multiples of $0.2$ T are shown for clarity. The blue curves correspond to the magnetic field values given in Fig.~\ref{fig:exp}(a). For all magnetic field values and over the full plotted voltage range repetitive behavior corresponding to Fabry-P\'{e}rot oscillations is visible. \label{fig:FullFabry}}
\end{figure}

One should note that, according to Ref.~\cite{Mireles2012}, the Zeeman splitting in BLG is around $1.1$ meV for $10$T. Using the conversion formula for top-gate voltages to energies from the Supplemental Materials of \cite{Kraft2018b} for the same device at slightly different voltages, the distance between $V_\mathrm{TG}=-12$\,V and $V_\mathrm{TG}=-8$\,V corresponds to the band splitting of $15.2$\,meV. Moreover, the density in the 2D region is not as low as in the constriction, since the split gates do not cover this region. Therefore, the total spin  polarization of these 2D bands cannot be achieved and, since we only observe faint oscillations on top of the plateaus, the finesse $F$ is small. As a result, this dependence of the conductance on the magnetic field is not experimentally resolved. These Fabry-P\'{e}rot oscillations are clearly visible in the differentiated differential conductance in Fig.~\ref{fig:FullFabry}, where they appear as small oscillations over the full top-gate and magnetic field range. 

\begin{figure}[t]
	\centering
	\includegraphics[width=.95\linewidth]{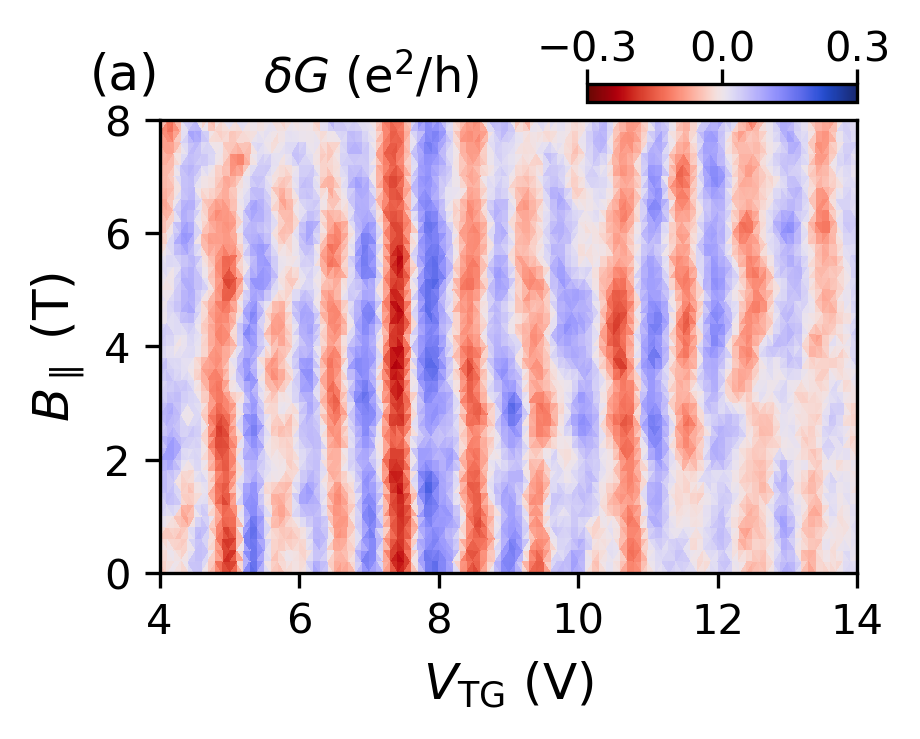}
	\includegraphics[width=.95\linewidth]{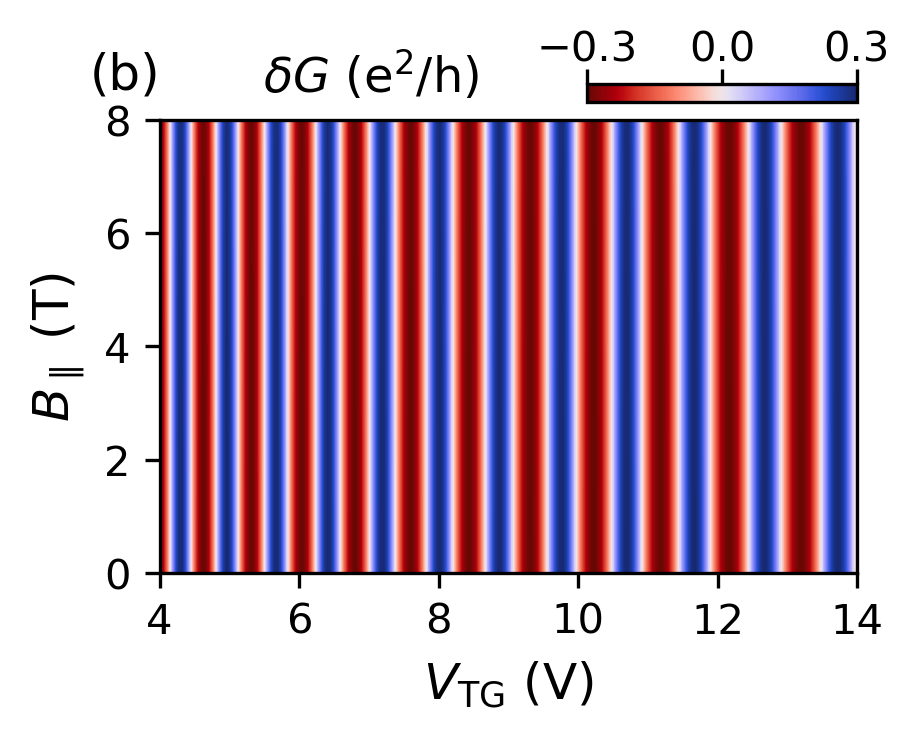}
\caption{(a) Oscillations of conductance $\delta G=G-\Bar{G}$for $V_\text{BG}=V_\text{SG}=0$. The smooth background conductance $\Bar{G}$ is obtained by means of a  Savitzky-Golay Filter of the measured data. (b) Fabry-P\'{e}rot oscillations given by Eq.~(\ref{eq:FP}) in a magnetic field. The Fermi vectors in magnetic field were obtained by calculating the gap from Eq.~(\ref{eq:asymmetry}) and the chemical potential, Eq.~(\ref{eq:chemPot}), for the given voltage and $n= c (V-V_0)$. The value $c=0.8\times 3.9\times 10^{15}$ was determined in Quantum Hall measurements and $V_0=-1.2$ V is the position of the Dirac point for $V_\text{BG}=V_\text{SG}=0$. The chosen length $L=230$\,nm corresponds to the distance between the leads and the split-gate fingers. The maxima and minima of oscillations stay nearly parallel over a large range of magnetic field.  \label{fig:Fabry}}
\end{figure}

An experimental example of the dependence of this conductance contribution on $V_\text{TG}$ and magnetic field is shown in Fig.~\ref{fig:Fabry}(a) for the case of vanishing back-gate and split-gate voltage and a theoretical plot based on Eq.~(\ref{eq:FP}) in Fig.~\ref{fig:Fabry}(b). The Fabry-P\'{e}rot resonances are seen for all magnetic field values and over the whole top-gate voltage range. These peaks, in contrast to the Zeeman-split subbands, only weakly depend on magnetic field, since the Zeeman splitting in the 2D bulk is smaller than in the QPC region. 

While the plots in Fig.~\ref{fig:Fabry} agree qualitatively, there are two points to keep in mind. The theoretical plot was obtained using no residual density, which is certainly not the case in the experiment, and does not account for the peculiarity of the screening in the experimental setup in the presence of the split gates. Even if the split-gate voltage is zero, the split gates affect the electrostatics of the setup by locally screening the top gate and developing mirror charges for carriers in this region. In addition, the dielectric layer in the split-gate region is noticeably thinner. This introduces inhomogeneity in the middle of the sample, and the length scale corresponding to the distance between the leads and the split gates naturally appears. 

In order to fully reproduce the experimentally observed pattern one would need a full electrostatic simulation. One very apparent difference is the fact that the period stays nearly constant in the experiment but not in the theory. While we did take into account the influence of the top-gate voltage on the density and the gap, the voltage also changes the boundary condition at the contacts and close to the split-gate fingers. Moreover we neglected any present residual density, which might change the position within the spectrum and thus the top-gate dependence. Finally, the presence of the split gates will naturally induce a breaking of translational symmetry, as the side boundaries do, which were not taken into account. For additional Fabry-P\'{e}rot interference data in the same sample, the reader is referred to the Supplemental Material of  Ref.~\cite{Kraft2018b}. A more thorough study of Fabry-P\'{e}rot interferences in BLG can be found in Ref.~\cite{Du2018}.

\section{Effective low-energy theory
\label{sec:eff2DModel}}
In order to derive the conductance of the system considered in the main text, we start with the effective two-band model for BLG, Eq.~(\ref{eq:BLGHam}). In this Appendix, we discuss how to obtain this low-energy approximation and how it is affected by the magnetic field, as well as by possible terms describing spin-orbit interaction.

Close to the $K$ point, the full four-band Hamiltonian is given by
\begin{align}
\mathcal{H}_0&=\begin{pmatrix}
\epsilon_{A1} & v\pi^{\dagger} & -v_4\pi^{\dagger} & v_3\pi ,\\
v\pi & \epsilon_{B1} & \gamma_1 & -v_4 \pi^{\dagger} ,\\
-v_4 \pi & \gamma_1 & \epsilon_{A2} & v\pi^{\dagger} ,\\
v_3\pi^{\dagger} & -v_4 \pi & v\pi & \epsilon_{B2}
\end{pmatrix} ,
\end{align}
where
\begin{align}
\epsilon_{A1}&=\frac{1}{2}(-U+\delta_{AB}),\\
\epsilon_{B1}&=\frac{1}{2}(-U+2\Delta^{\prime}-\delta_{AB}),\\
\epsilon_{A2}&=\frac{1}{2}(-U+2\Delta^{\prime}+\delta_{AB}),\\
\epsilon_{B2} &= \frac{1}{2}(U-\delta_{AB}).
\end{align}
It acts on the four-component wave-function according to
\begin{align}
\mathcal{H}\begin{pmatrix}
\psi_{A1}\\
\psi_{B1}\\
\psi_{A2}\\
\psi_{B2}
\end{pmatrix}&=E\begin{pmatrix}
\psi_{A1}\\
\psi_{B1}\\
\psi_{A2}\\
\psi_{B2}
\end{pmatrix}.
\end{align}
The spin degree of freedom is included through
\begin{align}
\mathcal{H}_i=\mathcal{H}_0\otimes \hat{s}_0
\end{align}
in the spin-degenerate sector, and the Zeeman splitting is introduced via
\begin{align}
\mathcal{H}_Z = \Delta E_Z\,\hat{\mu}_0\otimes\hat{\sigma}_0\otimes\hat{s}_z,
\end{align}
where $\mu,\sigma \text{ and }s$ are the Pauli matrices in layer, sublattice, and spin space, respectively. According to \cite{Guinea2010}, the two types of intrinsic spin-orbit interaction allowed by the symmetry of the problem are
\begin{align}
\mathcal{H}^\text{so}_1&=\xi\lambda_1\hat{\mu}_0\otimes\hat{\sigma}_z\otimes\hat{s}_z,\\
\mathcal{H}^\text{so}_2&=\xi\lambda_2\hat{\mu}_z\otimes\hat{\sigma}_0\otimes\hat{s}_z,
\end{align}
where $\xi=\pm$ corresponds to valley $K$ ($K'$).
Following \cite{Banszerus2020}, we additionally introduce an extrinsic spin-orbit interaction of the form
\begin{align}
\mathcal{H}^\text{so}_3=
\xi\,\mathrm{diag}(\lambda_u,-\lambda_u,\lambda_d,-\lambda_d)\otimes \hat{s}_z.
\end{align}

For the effective low-energy  theory, $|E|\ll\gamma_1$, we follow the derivation of Ref.~\cite{McCann2006a}. The basis states are reordered to $(\psi_{A1},\psi_{B2},\psi_{A2},\psi_{B1})\otimes(\uparrow,\downarrow)$, where the first half corresponds to the low-energy, non-dimer states and the second half to the dimer states, which are coupled by the large energy $\gamma_1$. From now on, all terms in the Hamiltonian are reordered according to this basis. Then, we can define the Greens function of the total Hamiltonian $\mathcal{H}=\mathcal{H}_i+\mathcal{H}_Z+\mathcal{H}^\text{so}_i$ as follows:
\begin{align}
\mathcal{H} & = \begin{pmatrix}
H_{11} & H_{12}\\
H_{21} & H_{22}
\end{pmatrix},\\
G & = \begin{pmatrix}
G_{11} & G_{12}\\
G_{21} & G_{22}
\end{pmatrix}=(\mathcal{H}-E)^{-1}
\notag
\\&
=\begin{pmatrix}
G_{11}^{(0)-1} & H_{12}\\
H_{21} & G_{22}^{(0)-1}
\end{pmatrix}^{-1},\\
G_{\alpha\alpha}^{(0)}
& =(H_{\alpha\alpha}-E)^{-1}.
\end{align}

The goal is to find a closed expression for $G_{11}$ which is then used to define the new Hamiltonian $\mathcal{H}_2$ according to
\begin{align}
G_{11}&=(\mathcal{H}_2-E)^{-1}\quad \Leftrightarrow 
\quad 
\mathcal{H}_2 = G_{11}^{-1}+E.
\end{align}
We find
\begin{align}
G_{11}=\left(1-G_{11}^{(0)}H_{12}G_{22}^{(0)}H_{21}\right)^{-1}G_{11}^{(0)}
\end{align}
and, thus,
\begin{align}
\mathcal{H}_2 = G_{11}^{-1}+E = H_{11}-H_{12}G_{22}^{(0)}H_{21}.
\end{align}
We now expand $G_{22}^{(0)}=(H_{22}-E)^{-1}$ in ${E/\gamma_1}\ll 1$.
Applying this procedure to the reordered full Hamiltonian produces, to linear order in $U,\Delta^{\prime}, \delta_{AB},v_4,v_3,\lambda,\Delta E_Z$, 
the effective two-band Hamiltonian:
\begin{align}
\mathcal{H}_2 = h_0+h_U+h_3+h_\text{AB}+h_{4}+h_{\Delta^{\prime}}+h_Z+h_\text{so},
\end{align}
where
\begin{align}
& h_0 = -\frac{v^2}{\gamma_1}
\begin{pmatrix}
0 & (\pi^{\dagger})^{2}\\
\pi^{2} & 0
\end{pmatrix}\otimes \hat{s}_0,
\\
& h_U = -\frac{U}{2}
\left[
\begin{pmatrix}
1 & 0\\
0 & -1
\end{pmatrix}-\frac{v^2}{\gamma_1^2}
\begin{pmatrix}
\pi^{\dagger}\pi & 0\\
0 & -\pi\pi^{\dagger}
\end{pmatrix}\right]\otimes \hat{s}_0,
\\
& h_3 = v_3
\begin{pmatrix}
0 & \pi\\
\pi^\dagger & 0
\end{pmatrix}\otimes \hat{s}_0,
\\
& h_\text{AB} = \frac{\delta_{AB}}{2}
\begin{pmatrix}
1 & 0\\
0 & -1
\end{pmatrix}\otimes \hat{s}_0,
\\
& h_{\Delta^{\prime}} 
= 2\Delta^{\prime} \frac{v^2}{\gamma_1^2}
\begin{pmatrix}
\pi^{\dagger}\pi & 0\\
0 & \pi\pi^{\dagger}
\end{pmatrix}\otimes \hat{s}_0,
\\
& h_{4} = 2v_4\frac{v^2}{\gamma_1^2}
\begin{pmatrix}
\pi^{\dagger}\pi & 0\\
0 & \pi\pi^{\dagger}
\end{pmatrix}\otimes \hat{s}_0,
\\
& h_Z = \frac{\Delta E_Z}{2}
\begin{pmatrix}
1 & 0\\
0 & 1
\end{pmatrix}\otimes \hat{s}_z,
\\
& h_\text{so} = \xi\begin{pmatrix}
\lambda_1+\lambda_2+\lambda_u & 0\\
0 & -\lambda_1-\lambda_2-\lambda_d
\end{pmatrix}\otimes \hat{s}_z.
\end{align}

\begin{figure}[ht!]
	\centering
	\includegraphics[width=0.95
\linewidth]{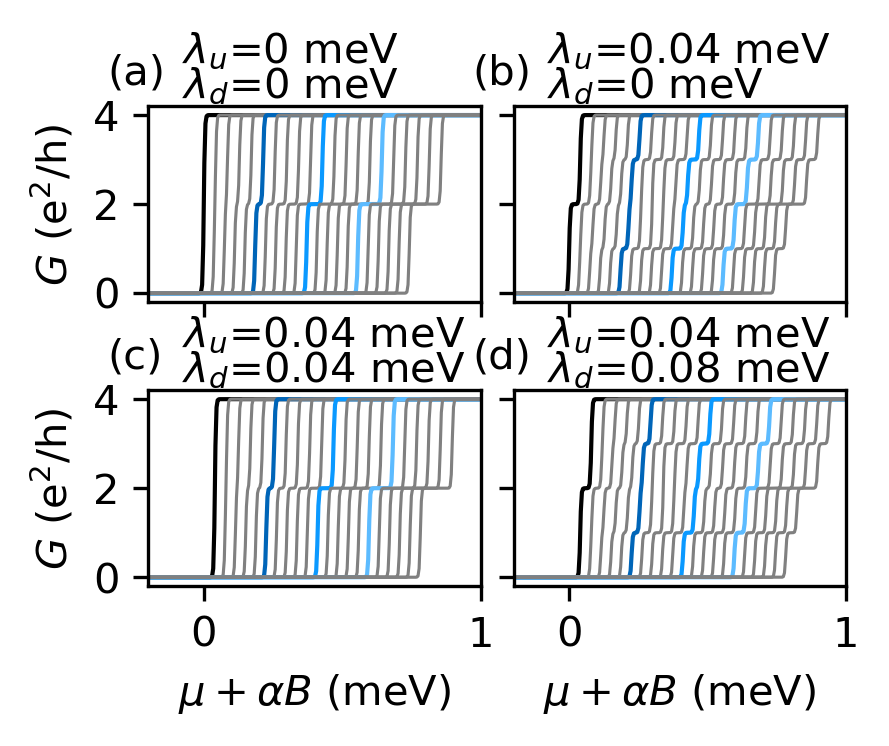}
	\caption{ Influence of a finite Kane-Mele spin-orbit coupling on the first conductance plateau in in-plane magnetic field, including the effect of layer-asymmetry encoded in $\lambda_u\neq \lambda_d$.  In all plots we chose $g=2$ and $\alpha=0.1$ meV/T, thus the rightmost curve corresponds to $B=8$~T and the temperature is $20$~mK. (a) No spin-orbit coupling, there is no splitting at zero magnetic field. (b) and (d) Different spin-orbit couplings in the two layers, finite splitting at zero magnetic field, full lifting of spin and valley degeneracy at higher magnetic field. (c) Same spin-orbit coupling in both layer, no splitting at zero magnetic field.  }
	\label{fig:SO}
\end{figure}

In the main text, we restrict ourselves to the terms $h_0$, $h_U$, and $h_Z$. 
This is exactly equation (\ref{eq:BLGHam}) and (\ref{eq:MexicanHam}). For all calculations, we furthermore neglect the second term of $h_U$
that describes the Mexican-hat feature of the spectrum. 
The only terms capable of lifting spin degeneracy are $h_Z$ and the spin-orbit term
\begin{align}
 \xi\begin{pmatrix}
\lambda_u & 0\\
0 & -\lambda_d
\end{pmatrix}\otimes \hat{s}_z
\end{align}
for asymmetry between the layers $\lambda_u\neq\lambda_d$, which can be caused by the lack of mirror symmetry of the whole stack \cite{Zollner2019}. 
Because of the valley index $\xi$ in this expression, the splitting is opposite in the two valleys, so that there is no net spin splitting due to spin-orbit interaction at all. If such a term is present in the Hamiltonian, it would lead to full spin-valley splitting in an applied magnetic field, i.e., four steps of $1\,e^2/h$.
This is, however, not seen in the experiment. This type of effect of spin-orbit coupling on the first conductance plateau in in-plane magnetic fields for the parameters specified in Ref.~\cite{Banszerus2020} is shown in Fig.~\ref{fig:SO}.

\section{Effect of tilted magnetic field}

We have investigated the effect of a perpendicular magnetic field on the quantized conductance in the same device in Ref.~\cite{Kraft2018b}. Large out-of-plane magnetic fields lead to a valley splitting, similar to the Zeeman spin splitting, with characteristic braiding behavior. Since we see neither a lifting of the valley degree of freedom, which would lead to a full resolution of conductance steps of $1\, e^2/h$ for large magnetic fields, nor any hint at a non-linear splitting, we can exclude a large out-of-plane component of the magnetic field. The presence of an appreciable out-of-plane component of the magnetic field would also show up in a curving of the Fabry-P\'{e}rot oscillations, which is also not observed here.

\begin{figure}[ht!]
	\centering
	\includegraphics[width=0.95
\linewidth]{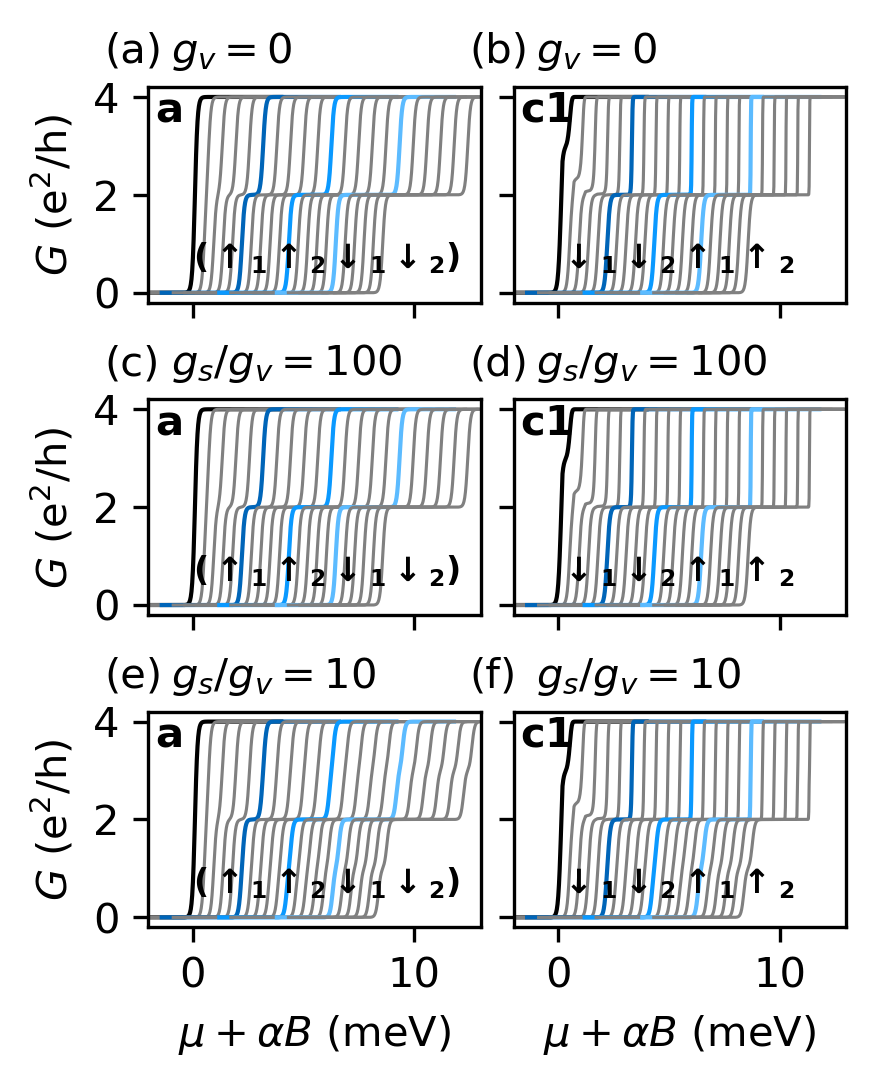}
	\caption{ Influence of finite tilt angle on the conductance traces for the cases $\mathbf{a}$ and $\mathbf{c1}$. In all plots, we chose $g_s=4$ and $\alpha=1.3$ meV/T, thus the rightmost curve corresponds to $B=8$~T and the temperature is $20$~mK. The configurations corresponding those in Fig.~\ref{fig:initial} are marked inside the panels; the index 1,2 corresponds to the $K,K^\prime$ valley respectively.  Panels (a) and (b): No finite tilt and thus no valley splitting. Panels (c) and (d): Finite tilt angle resulting in $g_v=0.04$. No change compared to $g_v$=0 is visible. Panels (e) and (f): Finite tilt angle resulting in $g_v=0.4$. At a certain magnetic field this would lead to the appearance of additional conductance steps at $1\, e^2/h$ and $3\,e^2/h$, which are not observed in our experiment.  }  
	\label{fig:Tilt}
\end{figure}

For small out-of-plane components the valley splitting is roughly linear \cite{Knothe2018} and can be easily included into our model by adding a term $\tau g_v B$ to the energy spectrum, where $\tau=\pm 1$ corresponds to the valleys $K$ and $K^\prime$, respectively, and $g_v$ contains both the angular dependence on the tilt angle and the magnetic moment due to the non-trivial Berry curvature. The expected effect of the tilt on the conductance traces is shown in Fig.~\ref{fig:Tilt}. We see that a very small tilt does not lead to any noticeable difference, while a bigger one leads to a full lifting of all degeneracies at strong magnetic fields.  
This is in contrast to quantum dots in BLG, see, e.g., Ref.~\cite{Eich2018}, where all four single-particle energies can be extracted at all values of magnetic field due to their additional charging energies and one can construct an effective $g$-factor by combining spin and valley splitting in a specific way, that would get enhanced over the bare spin Land\'{e} $g$-factor for one combination, while reducing it for the other combination. Since we do not observe a valley splitting, this effect would exactly average out in our case.

\section{QPCs in BLG}

In the experiment, we electrostatically induce a constriction in the shape of a QPC. While the full self-consistent treatment of this electrostatic problem is very involved, the main features can be described by the inclusion of a local potential in the corresponding Schr{\"o}dinger equation, thus neglecting the coupling between the Schr{\"o}dinger and the Poisson equations. There are two slightly different ways to include this potential profile. We start by summarizing the most common one for a normal two-dimensional Schr{\"o}dinger equation, and proceed with an alternative method for BLG. The obtained energy spectrum would extend dispersion (\ref{eq:BLGQuantDisp}) in the main text and thus also the density (\ref{eq:densitypermu}).

\subsection{Projection procedure for the Schrödinger equation \label{sec:projSch}}
In this section, we briefly recapitulate the projection procedure for a QPC in a conventional 2D electron gas, as discussed in 
Ref.~\cite{Buettiker1990}. The Hamiltonian is given by
\begin{align}
\mathcal{H}&=-\frac{\hbar^2}{2m}(\partial_x^2+\partial_y^2)+V(x,y),\\
V(x,y)&=V_\text{QPC}(x,y)+V_\text{lead}(x),\\
V_\text{QPC}(x,y) &= \frac{V_0}{\cosh(x/a)^2}+\frac{m \omega_y^2}{2}y^2,\\
V_\text{lead}(x)&=
V_l\left[\Theta(x-d)+\Theta(-d-x)\right],
\end{align}
where one models the QPC as a harmonic potential in the transverse direction and $V_l$ corresponds to the potential difference at the leads. 
We expand the wavefunction in transverse modes $\chi_{nx}(y)$ according to
\begin{align}
\Psi(x,y) = \sum_n \phi_n(x)\chi_{nx}(y),
\end{align}
where the complete orthonormal basis $\chi_{nx}(y)$ satisfies
\begin{align}
\left[-\frac{\hbar^2}{2m}\partial_y^2+V(x,y)
\right]\chi_{nx}(y)=\epsilon_n(x)\chi_{nx}(y).
\end{align}
For the given potential, these eigenvalues are given by 
\begin{align}
\epsilon_n(x)& = \hbar \omega_y \left(n+\frac{1}{2}\right)
+\frac{V_0}{\cosh(x/a)^2}
\notag
\\
&+V_l\left[\Theta(x-d)+\Theta(-d-x)\right]\nonumber
\end{align}
and from this we get the projected 1d problem 
\begin{align}
\left[ -\frac{\hbar^2}{2m}\partial_x^2+\epsilon_n(x)\right]\phi_n(x)=E\phi_n(x).
\end{align}
At low energies, only the lowest transverse modes contribute, and we can approximate the full solution as $\Psi(x,y)=\phi_0(x)\chi_{0x}(y)$. 
The exact form of the top of the effective 1D potential $\epsilon_n(x)$ determines at what position the additional $0.7$ shoulder appears; for the usual parabolic barrier top it appears very close to $0.7\times 2\,e^2/h$ \cite{Heyder2015}.
\subsection{Procedure for BLG \label{sec:projBLG}}
In the present case of BLG it is easier to  include the constriction by means of boundary conditions than a real local potential.  
The Hamiltonian\, (\ref{eq:BLGHam})  acts as 
\begin{align}
\hat{H}\begin{pmatrix}
\psi_{A1\uparrow}\\
\psi_{A1\downarrow}\\
\psi_{B2\uparrow}\\
\psi_{B2\downarrow}
\end{pmatrix}&=E\begin{pmatrix}
\psi_{A1\uparrow}\\
\psi_{A1\downarrow}\\
\psi_{B2\uparrow}\\
\psi_{B2\downarrow}
\end{pmatrix}
\end{align}
on the four-component wave-function in the spin and sublattice space. These four coupled second-order equations can be decoupled into a fourth-order one and we get for the first two components:  
\begin{align}
&\left[\left(\frac{U}{2}-U\frac{ v^2}{\gamma_1^2}\pi^{\dagger}\pi\right)^2+\frac{(\pi^{\dagger}\pi)^2}{(2m)^2}\right]\psi_{A1\sigma} \nonumber\\
&= \left(E-\sigma\frac{\Delta E_Z}{2}\right)^2\psi_{A1\sigma}.
\end{align}
Here, we have used that, without a magnetic field, the momentum operators commute. 


We expand the wave-function $\psi_{A1\sigma}$ in transverse modes $\chi_{nx\sigma}(y)$ as
\begin{align}
\psi_{A1\sigma}(x,y)&=\sum_n\phi_{n\sigma}(x)\chi_{nx\sigma}(y),
\end{align}
which leads to a new differential equation, where the $x$ and $y$ component are still coupled. We already assume, that $ \chi_{nx\sigma}(y)\propto
    \sin(k_{nx\sigma}y)[\cos(k_{nx\sigma}y)]$ if the solution is antisymmetric [symmetric] when describing the QPC by imposing hard-wall boundary conditions along the $y$ direction. The width of the channel $W(x)$ depends smoothly on $x$ and we get standing waves with wavevector $$k_{n}(x)=\frac{n\pi}{W(x)}.$$ We decouple the components by neglecting all $x$ derivatives of $\chi_{nx\sigma}(y)$, which leads to 
 the effective 1D equation: 
\begin{align}
&\left[\left(\frac{U}{2}+\hbar^2 U\frac{ v^2}{\gamma_1^2}[\partial_x^2-k_n^2(x)]\right)^2+\frac{\hbar^4[\partial_x^2-k_n^2(x)]^2}{(2m)^2}\right]\phi_{n\sigma}(x) \nonumber\\
&= (E-\sigma\frac{\Delta E_Z}{2})^2\phi_{n\sigma}(x),
\end{align}
where the constriction $W(x)$ acts as an effective 1D potential $E_n(x)=\frac{\hbar^2 n^2\pi^2}{2m W(x)^2}$. At low energies, only the lowest transverse mode contributes and we can approximate the full solution as $\psi_{A1\sigma}=\phi_{1\sigma}(x)\chi_{1x\sigma}(y).$ Choosing, for example, $W(x)=\cosh(x/L)$ we get a very realistic 1D potential containing terms of the form $1/\cosh(x/L)^2$.

\bibliography{Quellen}
\end{document}